 \documentstyle{article}
\input mssymb
\def\Box{\hbox{\vrule height1ex\kern-0.4pt
\vbox to 1ex{\hrule width1ex\vfil\hrule width1ex}\kern-0.4pt\vrule height1ex}}
\newcommand{\sqr}[2]{{{\vcenter{\vbox{\hrule height.#2pt
\hbox{\vrule width.#2pt height#1pt \kern#1pt
\vrule width.#2pt}
\hrule height.#2pt}}}}}

\topmargin = - 0.5 cm
\newcommand{\til}{\tilde}
\newtheorem{prop}{Proposition}
\newtheorem{theorem}{Theorem}
\newtheorem{coro}{Corollary}
\newtheorem{defn}{Definition}
\newtheorem{lemma}{Lemma}

\newcommand{\pco}{\pi_{\rm co}}

\newcommand{\cci}{C_c^{\infty}}
\newcommand{\cin}{C^{\infty}}

\newcommand{\eo}{\setcounter{equation}{0}}

\newcommand{\be}{\begin{equation}}
\newcommand{\bll}{\mbox{} \hfill \\}
\newcommand{\ee}{\end{equation}}

\newcommand{\al}{\alpha}

\newcommand{\Gm}{\Gamma}

\newcommand{\dl}{\delta}

\newcommand{\zt}{\zeta}

\newcommand{\th}{\theta}

\newcommand{\lm}{\lambda}

\newcommand{\rh}{\rho}
\newcommand{\sg}{\sigma}
\newcommand{\ta}{\tau}
\newcommand{\ph}{\phi}
\newcommand{\phv}{\varphi}
\newcommand{\ch}{\chi}
\newcommand{\ovl}{\overline}

\newcommand{\ps}{\psi}
\newcommand{\om}{\omega}
\newcommand{\Om}{\Omega}

\newcommand{\raw}{\rightarrow}
\newcommand{\law}{\leftarrow}

\newcommand{\A}{{\cal A}}
\renewcommand{\L}{{\cal L}}
\newcommand{\B}{{\cal B}}

\newcommand{\F}{{\cal F}}
\newcommand{\G}{{\cal G}}

\newcommand{\bib}{\bibitem}

\renewcommand{\H}{{\cal H}}

\newcommand{\pot}{\stackrel{\circ}{p}}

\newcommand{\rab}{\rangle_{\cal B}}

\newcommand{\g}{\mbox{\bf g}}

\newcommand{\h}{\mbox{\bf h}}

\newcommand{\n}{\parallel}

 \renewcommand{\ll}{\label}
 \newcommand{\fn}{\footnote}
\newcommand{\plg}{\pi_{\chi}}
\newcommand{\pug}{\pi^{\chi}}

\renewcommand{\O}{{\cal O}}
\newcommand{\K}{{\cal K}}

\newcommand{\bone}{{\mbox{{\bf 1}}}}

\newcommand{\hlg}{{\cal H}_{\chi}}
\newcommand{\hug}{{\cal H}^{\chi}}

\newcommand{\pl}{\hbar}
\newcommand{\notp}{p \kern-.48em /}
\newcommand{\ci}{\cite}

\newcommand{\ot}{\otimes}

\newcommand{\bea}{\begin{eqnarray}}
\newcommand{\eea}{\end{eqnarray}}
\newcommand{\la}{\langle}
\newcommand{\ra}{\rangle}
\newcommand{\RE}{\Bbb R}
\begin{document}
\setlength{\baselineskip}{1\baselineskip}
\thispagestyle{empty}
\title{Rieffel induction as generalized quantum Marsden-Weinstein reduction}
\author{
N.P.~Landsman\thanks{ Supported by an S.E.R.C. Advanced Research Fellowship}\\
\mbox{}\hfill \\
Department of Applied Mathematics and Theoretical Physics\\
University of Cambridge, Silver Street\\
Cambridge CB3 9EW, United Kingdom \\ \bll Preprint DAMTP-93-22}
\maketitle
\begin{abstract}
A new approach to the quantization of constrained or otherwise reduced
classical mechanical systems
is proposed. On the classical side, the generalized symplectic reduction
procedure of Mikami and
Weinstein, as further extended by Xu in connection with symplectic equivalence
bimodules and Morita
equivalence of Poisson manifolds, is rewritten so as to avoid the use of
symplectic groupoids, whose
quantum analogue is unknown. A theorem on symplectic reduction in stages is
given. This allows one to
discern that the `quantization' of the generalized  moment map consists of an
operator-valued inner
product on a (pre-) Hilbert space (that is, a structure similar to a Hilbert
$C^*$-module). Hence
Rieffel's far-reaching operator-algebraic generalization of the notion of an
induced representation
is seen to be the exact quantum counterpart of the classical idea of symplectic
reduction, with
imprimitivity bimodules and strong Morita equivalence of $C^*$-algebras falling
in the right place.

Various examples involving groups as well as groupoids are given, and known
difficulties with both
Dirac and BRST quantization are seen to be absent in our approach.
 \end{abstract}
 \newpage
\section{Introduction}
Marsden-Weinstein reduction \ci{MaW,Mey} (alternatively known as Hamiltonian or
symplectic reduction)
plays a basic role in classical mechanics \ci{AM,GS,LM,Mar}, as well as in pure
mathematics.
 The starting point is a connected symplectic manifold $S$ equipped with an
right-action
of a Lie group $H$ (assumed connected for simplicity), which action we assume
to be strongly
Hamiltonian for the moment. In that case one has an equivariant moment map
$J:S\raw (\h^*)^-$, where
$\h$ is the Lie algebra of $H$, and $\h^*$ its dual (that is, $J$ intertwines
the co-adjoint action
on $\h^*$ and the action on $S$); here and in what follows the notation $P^-$
stands for a Poisson
manifold $P$, equipped with minus its original Poisson structure.  The
essential point is that the
pull-back $J^*:\cin((\h^*)^-)\raw \cin(S)$ is a morphism of Poisson algebras
(relative to the
Lie-Poisson structure on $\h^*$ \ci{GS,Wei83,LM,MiW}). The choice of a
co-adjoint orbit $\O\in\h^*$
then leads to the reduced space $S^{\O}=J^{-1}(\O)/H$, which inherits a
symplectic structure from
$S$. If $A$ is a Poisson subalgebra of $\cin(S)$ whose elements are
$H$-invariant (equivalently, they
Poisson-commute\fn{This condition is not strictly necessary, but facilitates
the presentation,
 and is satisfied in generic examples} with $J^* \cin(\h^*)$), we obtain a
Poisson morphism
$\pi^{\O}:A\raw \cin(S^{\O})$. This may be thought of as a `classical
representation' of $A$ on
$S^{\O}$, which is induced from the Poisson morphism (or, once again,
`classical representation')
$\pi_{\O}\equiv i_{\O}^*: \cin(\h^*)\raw \cin(\O)$, where $ i_{\O}$ is the
inclusion map of $\O$ into
$\h^*$. For example, one is usually given  an $H$-invariant Hamiltonian
$H_0\in\cin(S)$, whose
representative $\pi^{\O}(H)\in\cin(S^{\O})$ is the reduced Hamiltonian on the
reduced phase
space. More generally, any symplectic realization $\rh: X\raw \h^*$ of the
Poisson manifold $\h^*$
(that is,  $X$ is symplectic, and $\rh$ is a Poisson map \ci{CDW}) leads to a
classical
representation $\pi^X(A)$ on a certain symplectic space $S^X$, to be detailed
below.

The connection with constrained mechanical systems \`{a}  la Dirac \ci{Dir} is
as follows: one chooses
a basis $\{T_i\}_{i=1,\ldots d_H}$ of $\h$, and defines $f_i\in\cin(S)$ by
$f_i=J^*\til{T}_i$; here
$\til{T}_i\in\cin(\h^*)$ is defined by $\til{T}_i(\th)=\la\th,T_i\ra$ for
$\th\in\h^*$. Then pick an
arbitrary point $\mu\in\O$, put $\mu_i=\til{T}_i(\mu)=\la \mu,T_i\ra$, and take
the constraints on $S$
to be $\Phi=f_i-\mu_i=0,\, i=1\ldots, d_H$. These constraints will in general
be  mixed (that is, of
first as well as second class), and one obtains the reduced phase space by
quotienting the
constraint surface by the foliation defined by the Hamiltonian flows of the
first-class constraints
\ci{Dir}. This reduced phase space of Dirac is then symplectomorphic to the
Marsden-Weinstein reduced
space $S^{\O}$ mentioned above. The geometric procedure is superior to the
`physicists' approach just
sketched, in that one need not pick a basis of $\h$,  an arbitrary point $\mu$,
or explicitly
classify the functions of constraint $\Phi_i$ into first and second class ones.

 One would naturally like to generalize this construction to the situation
where one has a symplectic
space $S$, a Poisson manifold $P$ \ci{LM}, and two Poisson maps $J:S\raw P^-$
and $\rh:X\raw P$,
where $X$ is symplectic. This should lead to an `induced classical
representation' $\pi^X$ of any
Poisson subalgebra $A\subset \cin(S)$ which Poisson-commutes with $J^*\cin(P)$,
on some symplectic
space $S^X$. This generalization was partly achieved by Mikami and Weinstein
\ci{MiW} in the special
case where $P$ is integrable (in the sense that it is the base space of units
of a symplectic groupoid
\ci{CDW,MiW}), and $X$ is a symplectic leaf of $P$ (with $\rh$ the injection
map), and later Xu
\ci{Xu89} gave a more general construction avoiding the latter restriction. A
slight rewriting of
this, finally lifting also the condition that $P$ be integrable (and thereby
avoiding constructions
involving symplectic groupoids, whose quantization we do not understand), is
given in section 2
below.

{}From the physical point of view of constrained systems, what this
generalization achieves is that
now reduced phase spaces obtained from arbitrary Poisson algebras of
constraints may be described in
a very satisfactory geometric fashion. The physicist's approach would be to
choose a basis $\til{T}_i$
which generates $\cin(P)$ in some appropriate way, and pick a point $\mu\in L$,
where $L$ is a
symplectic leaf in $P$. With the $f_i$ (which satisfy the Poisson algebra of
$P^-$) and $\mu_i$
defined as above, one then easily finds that the reduced phase space defined by
the constraints
$\Phi_i$ is symplectomorphic to $S^L$. However, if $X$ in the preceding
paragraph is not taken as
a symplectic leaf in $P$, one obtains a symplectic space $S^X$ (and an
associated representation of
the Poisson algebra $A$) which cannot be obtained as a reduced phase space in
the traditional sense,
in any obvious way.

Thus one  has a very general method of constructing new symplectic spaces and
Poisson morphisms from
old ones at one's disposal, which ought to be quantized in some way. While a
direct quantization of
the reduced symplectic manifolds and concordant induced representations of
Poisson algebras may be
possible in certain examples, a systematic approach intending to mimic the
classical
reduction/induction procedure in some quantum fashion ought to start from a
quantization of the
`unconstrained' system. Hence we assume we have found two commuting operator
algebras $\A$ and
$\B$ acting on some Hilbert space $\H$ from the left and from the right,
respectively, as well as a
(left) $\B$-module $\H_{\ch}$; from these data we try to construct an `induced'
representation
$\pi^{\ch}(\A)$ on a Hilbert space $\hug$. We denote these data by $\A\raw \H
\law \B$ and $\B
\stackrel{\plg}{\raw}\hlg$.

For our purpose it does not matter very much what one
exactly means by a quantization; the induction procedure may be applied to any
data
$\H,\A,\B,\plg,\hlg$. Ideally, these data correspond to a strict deformation
quantization
\ci{Rie89b} (as redefined in \ci{NPLstr}) of the symplectic data, as in some of
our examples in
section 4.

We now take our cue from  three directions (details to be given later on in
this paper):
\begin{enumerate}
\item
Take $G$ a locally compact group and  $H\subset G$ a closed subgroup. Let
$\plg(H)$ be a unitary
representation of $H$ on $\hlg$; we may then form the induced representation
$\pug(G)$ on a specific
Hilbert space $\hug$, as defined in the Mackey theory \ci{Mac,Var}. Rieffel
\ci{Rie74} relates this to
the data $C^*(G)\raw L^2(G)\law C^*(H)$  and $C^*(H)
\stackrel{\plg}{\raw}\hlg$, where $C^*(G)$ is the group algebra of $G$
\ci{Ped}, which acts on $L^2(G)$ in the left-regular representation, with
$C^*(H)$ acting in the
right-regular anti-representation (restricted to $H$).

 In case that $G$ and $H$ are a Lie groups, it is
argued in \ci{KKS,GS,Wei87} that the classical analogue of the Mackey induction
procedure is to take
$S=T^*G$, $A=\cin(\g^*)\simeq \cin(T^*G)^G$ (the Poisson algebra of smooth
functions on $T^*G$
which commute with the pull-back to $T^*G$ of the right-action of $G$ on
itself), and $P=\h^*$; the
moment map $J:T^*G\raw (\h^*)^-$ comes from the pull-back of the right-action
of $H$ on $G$. A
co-adjoint orbit $\O\subset \h^*$ is then analogous to an  irreducible unitary
representation $\plg$,
and the Marsden-Weinstein reduced space $J^{-1}(\O)/H\equiv (T^*G)^{\O}$,
carrying the induced action
$\pi^{\O}$ of $\cin(\g^*)$, is the symplectic counterpart of the Hilbert space
$\hug$ carrying the
induced representation $\pug$ of $G$ (or $C^*(G)$). To complete the parallel,
we recall Rieffel's
discovery that the group algebra $C^*(G)$ is a  deformation quantization  of
the
Poisson algebra $\cin(\g^*)$ \ci{Rie89a}, which in specific cases is even
strict in the sense
of \ci{Rie89b}.
\item
Let $(P,Q,H,pr)$ be a principal fibre bundle  with projection $pr:P\raw Q$ and
a compact Lie group $H$
acting on the total space $P$ from the right. The symplectic leaves  of the
Poisson manifold
$P=(T^*P)/H$ are in one-to-one correspondence with the co-adjoint orbits $\O$
in $\h^*$, and, as
originally discovered by Sternberg,  a leaf
 $S^{\O}$ plays the role of the phase space of a particle moving on $Q$ with
internal charge $\O$,
which couples to a Yang-Mills field with gauge group $H$ \ci{GS,Wei78,Mar}.
This is evidently described through Marsden-Weinstein reduction by taking
$S=T^*P$ and
$A=\cin(T^*P)^H\simeq \cin((T^*P)/H)$.

The quantization of this setting was constructed in \ci{NPLstr} using some Lie
groupoid and algebroid
technology. The results were obtained by applying a generalized induction
procedure to the
quantum data $\K(L^2(P))^H\raw L^2(P) \law C^*(H)$ and $C^*(H)
\stackrel{\plg}{\raw}\hlg$, thus
obtaining irreducible representations of the $C^*$-algebra $\K(L^2(P))^H$ of
$H$-invariant compact
operators on $L^2(P)$ on spaces $\hug$ analogous to the one used in the Mackey
theory. Indeed, the
special case $P=G$ reproduces the constructions in the previous item. One
obtains a sharpened
version of a strict deformation quantization even in the general case. A simple
special case of this
construction appeared in \ci{NPLIR}.
 \item
It was recognized by Xu \ci{Xu91} that a complete full dual pair
$P_1\stackrel{J_1}{\law} S
\stackrel{J_2}{\raw} P_2^-$  of Poisson
manifolds \ci{Wei83} (with connected and simply connected fibers) defines an
equivalence bimodule of
the corresponding Poisson algebras. Hence there is a bijective correspondence
between the categories
of symplectic realizations of $P_1$ and $P_2$, respectively \ci{Xu89}; from an
algebraic point of view
this means that the Poisson algebras $\cin(P_1)$ and $\cin(P_2)$ have
equivalent classical
representation theories. This equivalence is implemented through a generalized
symplectic reduction
procedure (see subsect.\ 2.1 below).

There is an obvious formal analogy between these classical equivalence
bimodules, and
the imprimitivity bimodules $\A\raw \H \law \B$ of operator algebras defined by
Rieffel \ci{Rie74}.
Under certain conditions, the main one being the existence of compatible
rigging maps
$\langle\,,\,\rangle_{\B}:\H\times \H\raw \B$ and
$\mbox{}_{\A}\langle\,,\,\rangle:\H\times \H\raw \A$,
the representation theories of $\A$ and $\B$ are isomorphic, and the
isomorphism is implemented by a
generalized induction procedure given in \ci{Rie74} called {\em Rieffel
induction} \ci{FD}.

Indeed, the term `Morita equivalence of Poisson manifolds' \ci{Xu91} was
clearly inspired by the
terminology of (strong) `Morita equivalence of operator algebras'
\ci{Rie74,Rie82}.
For example, under certain conditions (cf.\ subsect.\ 4.3) the Poisson manifold
$(T^*P)/H$ is Morita
equivalent to $\h^*$ through the equivalence bimodule $T^*P$, and on the
quantum side we find strong
Morita equivalence of the $C^*$-algebras $\K(L^2(P))^H$ and $C^*(H)$ through
the imprimitivity
bimodule $L^2(P)$.
 \end{enumerate}
In the light of the above evidence, and more to be given in the main body of
the paper, it is not
very daring to suggest that the quantum analogue of the generalized symplectic
reduction procedure
sketched earlier, is provided by Rieffel induction. We will now briefly
describe this
construction (cf.\ \ci{Rie74,FD} for an exhaustive treatment, or sect.\  3.1
below for a brief
summary of rigging maps and Rieffel induction).

In symplectic geometry, a Poisson map $J:S\raw P^-$
plays a double role: it relates $S$ to $P$, {\em and} provides a Poisson
morphism
$J^*:\cin(P^-)\raw \cin(S)$. In operator theory, a (right) action of a
$\mbox{}^*$-algebra $\B$ on a
Hilbert space $\H$ amounts to a $\mbox{}^*$-anti-homomorphism $\pi^-:\B\raw
{\cal L}(\H)$, which is
the `quantum' analogue of $J^*$.  It is now tempting to define a quantum
version of $J$ as some map
between the projective space $\Bbb P \H$ and the state space of $\B$, and
construct an induction
procedure on this basis, but this appears to lead nowhere unless $\B=C^*(H)$
for compact $H$; the
correct `quantization' of the moment map is a so-called rigging map
(alternatively called an
operator-valued inner product).

 In general, this is a map $\la\cdot,\cdot\rab$ defined  on $L\otimes L$
(algebraic tensor product),
where $L\subset \H$, taking values in $\B$, for which $\la\ps,\phv
B\rab=\la\ps,\phv\rab B$ for all
$\ps,\phv\in L$ and all $B\in\B$.  In case that $\B=C^*(H)$, so that $\pi^-$
above is defined through
a unitary representation $\pi$ of $H$ on $\H$, the rigging map is defined by
$\la\ps,\phv\ra_{C^*(H)}:h\raw (\pi(h)\phv,\ps)$, where $(\cdot ,\cdot)$ is the
inner product on $\H$.
This defines a function $f_{\ps,\phv}$ on $H$, and we choose $L$ in such a way
that  $f_{\ps,\phv}\in
C^*(H)$ for all $\ps,\, \phv\in L$. If $H$ is compact we can simply take
$L=\H$. In the non-compact
case, for e.g., $\H=L^2(G)$ with $H$ acting on the right, one may take
$L=C_c(G)$. As we will see, it
is easier in practice to start with the dense subalgebra $C_c\subset C^*(H)$ in
the above
consideration.

Now suppose that another $\mbox{}^*$-algebra $\A$ acts on $L$, and the
condition $\la
A\ps,\phv\ra_{\B}=\la \ps, A^*\phv\ra_{\B}$ is satisfied for all $A\in \A$
(this is certainly the case
if $\A$ commutes with $\B$). Under favourable circumstances a representation
$\plg(\B)$ on a Hilbert
space $\hlg$ may then be induced to a representation $\pug(\A)$ on a certain
Hilbert space $\hug$. The
crucial step in this induction procedure is to start with $L\otimes \hlg$,
equipped with a
sesquilinear form $(\cdot ,\cdot)_0$ defined by $(\ps\ot v,\phv\ot
w)_0=(\plg(\la
\phv,\ps\ra_{\B})v,w)_{\hlg}$; this is positive semi-definite if the rigging
map is positive, and in
that case one may quotient $L\ot\H$ by the null space of $(\cdot ,\cdot)_0$,
and complete it into a
Hilbert space $\hug$, which inherits the left-action of $\A$ from $L$.

Forming $L\ot\H$ with the given
sesquilinear form is the quantum counterpart of taking  $J^{-1}(\mu)\subset S$
(for some $\mu\in\O$)
with its pre-symplectic form borrowed from $S$ in the Marsden-Weinstein
reduction process, and
quotienting the null space of $(\cdot ,\cdot)_0$ away is obviously the quantum
analogue of quotienting
$J^{-1}(\mu)$ by its characteristic (null) foliation, thus obtaining a
symplectic space
symplectomorphic to $J^{-1}(\O)/H$. These formal analogies will be more clearly
visible in the
description of the generalized symplectic reduction procedure defined in
subsect.\ 2.1 below.

A point $\mu$ may fail to be a regular value of the moment map \ci{AM,Mar},
which leads to some
difficulties in the reduction procedure. This problematic situation is
`quantized' by the potential
existence of vectors $\ps\in L$ for which $\plg(\la \ps,\ps\ra_{\B})$ fails to
be a positive operator
on $\hlg$, so that $(\, ,\,)_0$ is not positive semi-definite. Evidently, this
problem will not arise
if the rigging map is positive. In general, the quantum reduction procedure is
better behaved than
its classical counterpart, cf.\ Prop.\ 4.

In the remainder of this paper we will describe the above ideas in detail, and
provide a fair number
of examples illustrating why it seems a good idea to quantize the generalized
symplectic
reduction/induction technique by the Rieffel induction process. For example, we
give classical
Poisson versions of both the imprimitivity theorem and the theorem on induction
in stages
\ci{Rie74,FD}.

 The quantization procedure based on Rieffel induction will have to be compared
with the fashionable
BRST quantization scheme (cf.\ e.g.\ \ci{KS,Hue}). For the moment, we just wish
to point out that
serious difficulties of principle with the latter were spelled out in
\ci{DET,LL},
and that on the practical side ``at present the computation of BRST-cohomology
is an extremely
difficult problem'' \ci{Hue}. Moreover, the Rieffel induction process mimics
the symplectic
procedure  more closely than any BRST treatment we are aware of (including the
bosonic BRST theory in
\ci{DEGST}), and appears to be
 simpler both conceptually and computationally. On the other hand, the proper
domain of the rigging
map has to be found case by case, and for $C^*$-algebras not defined by
groupoids even the rigging map
itself is not given {\em a priori}.  Finally, all our examples are defined for
finite-dimensional
Poisson manifolds, and one has yet to see how quantization through Rieffel
induction will perform in
infinite-dimensional situations (where the BRST technique has been very
successful \ci{KS}).
\section{Symplectic induction}\eo
\subsection{Generalized Marsden-Weinstein reduction}
As pointed out in the Introduction,   Marsden-Weinstein reduction is  a special
case of a more
general symplectic  induction technique. The general procedure described below
is essentialy due to Xu
\ci[Prop.\ 2.1]{Xu89}. By rewriting his construction omitting any reference to
symplectic groupoids,
we are able to avoid the restriction in \ci{Xu89} to integrable Poisson
manifolds (in the sense of
\ci{CDW,MiW}), while also the parallel with the Rieffel induction technique in
the quantum case is
more transparent in this way. \begin{defn} Let $S$ and $S_{\rh}$ be connected
symplectic manifolds,
$P$ a Poisson manifold, $P^-$ the same manifold as $P$ but equipped with minus
its Poisson bracket,
and let $J:S\raw P^-$ and $\rh:S_{\rh}\raw P$ be Poisson maps. Then $S*_{P}
S_{\rh}\subset S\times
S_{\rh}$ is defined by \be
 S*_{P} S_{\rh} =\{(x,y)\in S\times S_{\rh}| J(x)=\rh(y)\}. \ll{2.1}
\ee
Each $f\in\cin(P)$ defines a vector field $\hat{X}_f$ on $S\times S_{\rh}$ by
\be
\hat{X}_f\, g=\{J^*f-\rh^*f,g\}, \ll{2.2}
\ee
where the Poisson bracket is the product one on $S\times S_{\rh}$. \ll{def1}
\end{defn}
\begin{theorem}
  $S*_{P} S_{\rh}$ is
co-isotropically immersed in $S\times S_{\rh}$. The collection of vector fields
$\{ \hat{X}_f|
f\in\cin(P)\}$ defines a (generally singular) foliation ${\cal F}$, of  $S*_{P}
S_{\rh}$, whose leaf
space $S^{\rh}=S*_{P} S_{\rh}/{\cal F}$ coincides with the quotient of
 $S*_{P} S_{\rh}$ by its characteristic (null) foliation. \ll{cis}
\end{theorem}
{\em Proof}. The dimension counting argument in the proof comes from \ci{KKS}
and \ci{Xu89}.

We write $M$ for $S*_{P} S_{\rh}$ for simplicity. Let $X\in T_xS$ and $Y\in
T_yS_{\rh}$; then $X+Y\in
T_{(x,y)}M$ iff $J_*X=\rh_*Y$. The dimension of $T_{(x,y)}M$ at any point
$(x,y)\in M$ equals $\dim S+\dim S_{\rh} -({\rm rank}\, J_*)(x)$, so that the
dimension of
$T_{(x,y)}M^{\perp}$ (the symplectic orthoplement of $TM$ in $T(S\times
S_{\rh})$ at $(x,y)$) is
$({\rm rank}\, J_*)(x)$. Let ${\cal F}_{(x,y)}$ denote the linear span of the
collection of vector fields
$\hat{X}_f$ taken at $(x,y)$, where $f$ runs through $\cin(P)$. Then $\dim
{\cal F}_{(x,y)}=({\rm rank}\, J_*)(x)$. We next show that ${\cal F}_{(x,y)}
\subseteq T_{(x,y)}M^{\perp}$, so that
in fact  ${\cal F}_{(x,y)}= T_{(x,y)}M^{\perp}$. Namely, let $X+Y\in
T_{(x,y)}M$, as above; then with
$\om=\om_S+\om_{S_{\rh}}$ the symplectic form on $S\times S_{\rh}$, one has
 $$\la \om|X+Y,\hat{X}_f\ra_{(x,y)}=\la d(J^*f-\rh^*f)|X+Y\ra_{(x,y)} =0.$$
Moreover,  ${\cal F}_{(x,y)}\subset T_{(x,y)}M$ by a similar calculation: if
$X_g$ is the
Hamiltonian vector field of $g$, then by Lemma 1.2 in \ci{Wei83} $J_*
X_{J^*f}=-X_f$, where $X_f$ is defined w.r.t. the Poisson bracket on $P$
(rather than $P^-$, hence
the sign) and $\rh_*X_{\rh^*f}=X_f$. Thus $\hat{X}_f=X_{J^*f}-X_{\rh^*f}\in
TM$.
 Therefore, $M$ is
co-isotropically immersed in $S\times S_{\rh}$.
Furthermore, $[\hat{X}_f,\hat{X}_g]=-\hat{X}_{\{f,g\}}$ (Poisson bracket on
$P$), so that by the
Stefan-Sussmann theorems (cf.\ \ci[Thm.\ 3.9, 3.10., App.\ 3]{LM}) the
distribution $\cal F$ defines a
(singular) foliation, called $\cal F$ as well.  \Box

Under the additional assumption that $S^{\rh}$ is a manifold, we have
accordingly found a new symplectic
space $S^{\rh}$, which carries a `classical representation' of certain Poisson
subalgebras of $\cin(S)$,
as follows. We borrow some notation from operator algebras: if $B$ is a subset
of $\cin(S)$, then $B'$
denotes its Poisson commutant, i.e. the set of all functions in $\cin(S)$ whose
Poisson bracket with
each element of $B$ vanishes. Also, $[x,y]\in S^{\rh}\equiv (S\times_P
S_{\rh})/{\cal F}$ stands for the
equivalence class of a point $(x,y)\in S\times_P S_{\rh}$ under the foliation
$\cal F$.
 \begin{prop}
Let $A\subseteq (J^*\cin(P))' \subset \cin(S)$. Then the map $\pi^{\rh}:A\raw
\cin(S^{\rh})$, defined
by
\be \pi^{\rh}(f)([x,y])=f(x) \ll{2.a}
\ee
 is well-defined, and is a Poisson morphism. \ll{cir}
\end{prop}
This is obvious. We call $\pi^{\rh}$ the classical representation of $A$
induced by the map
$\rh:S_{\rh}\raw P$.
Suppose that we have a Poisson manifold $P_2$, and a Poisson map $J_2:S\raw
P_2$, such that
$J_2^*\cin(P_2)\subset \cin(S)$ Poisson-commutes with $J^*\cin(P)$; then the
proposition is
equivalent to the production of a Poisson map $J^{\rh}:S^{\rh}\raw P_2$ defined
by $J^{\rh}([x,y])=J_2(x)$.

It may be worth spelling out how Marsden-Weinstein reduction emerges as a
special case.
We take a connected Lie group $H$ acting on $S$ from the right in a strongly
Hamiltonian fashion
\ci{GS,LM} (see subsect.\ 3.3 below for the general case), so that there is an
equivariant moment
map $J:S\raw (\h^*)^-$ (hence $P=\h^*$).  (If a left-action with moment map
$J^-:S\raw \h^*$ is given,
simply put $J=-J^-$.) We then take $S_{\rh}=\O$, a co-adjoint orbit in $\h^*$,
and $\rh:\O\raw \h^*$
to be the inclusion map, which is evidently a Poisson map if $\h^*$ and $\O$
are endowed with the
Lie-Poisson structure. Then clearly $(S\times_{\h^*} \O)/{\cal F}\simeq
J^{-1}(\O)/H$, but note that
the null foliation of $J^{-1}(\O)\subset S$ does not coincide with the
$H$-foliation (whereas these
foliations do coincide on $S\times_{\h^*} \O\subset S\times \O$). Hence the
above diffeomorphism is
an efficient way of providing $J^{-1}(\O)/H$ with its correct symplectic
structure (which is usually
obtained from the diffeomorphism $J^{-1}(\O)/H\simeq J^{-1}(\mu)/H_{\mu}$,
where $\mu\in\O$ is
arbitrary, and $H_{\mu}$ is its stabilizer).

Another special case is the Mikami-Weinstein reduction procedure \ci{MiW}. They
assume that $P$,
which is $\Gamma_0$ in their notation, is the unit space of a symplectic
groupoid, and their reduced
space $J^{-1}(u)/\Gm_u$ emerges from Theorem 1 by taking $S_{\rh}$ to be the
symplectic leaf in $\Gm_0$
containing $u$, and $J_1$ the inclusion map in $\Gm_0$.
 \subsection{Symplectic imprimitivity theorem}
The well-known imprimitivity theorem of Mackey \ci{Var} has a far-reaching
generalization due to
Rieffel \ci{Rie74,FD,Rie82}. This generalization establishes a bijective
correspondence between  the
respective representation theories of two operator algebras satisfying a
certain equivalence
relation, known as strong Morita equivalence.  A satisfactory `classical' (that
is, Poisson-algebraic)
analogue of this equivalence relation and some of its ramifications was
recently given by Xu
\ci{Xu91}. For the convenience of the reader, we repeat Xu's definition of
Morita equivalent Poisson
manifolds (Def.\ 2.1 in \ci{Xu91}; the concept of  dual pair, which is central
to the definition, is
due to Weinstein \ci{Wei83}).
\begin{defn}
 A classical equivalence bimodule of a pair of Poisson manifolds $(P_1,P_2)$
consists of  a
symplectic manifold $S$ and a pair of Poisson morphisms $J_1:S\raw P_1^-$ and
$J_2:S\raw P_2$, such
that $P_2 \stackrel{J_2}{\law} S \stackrel{J_1}{\raw} P_1^{-}$ is a complete
full dual pair with
connected and simply connected fibers. This means that $J^*_1\cin(P_1^{-})$ and
$J^*_2 \cin(P_2)$ are
each other's Poisson commutant in $\cin(S)$, that the leaf spaces of the
foliations defined by the
fibers of $J_1$ and $J_2$ are manifolds in the quotient topology, and that
$J_1$ and $J_2$ are
surjective, as well as complete as Poisson maps.

Poisson manifolds $P_1$ and $P_2$ are called Morita equivalent if there exists
a classical
equivalence bimodule in the above sense. \ll{ceb}
\end{defn}
A Poisson map $J:S\raw P$ is said to be complete if  the  Hamiltonian vector
field $X_{J^*f}$ on $S$
is complete (that is, has a flow defined for all times) if $X_f$ on $P$ is, for
all $f\in\cin(P)$.
This condition is the classical analogue of the requirement that a
representation of a
$\mbox{}^*$-algebra on a Hilbert space be $\mbox{}^*$-preserving, that is, it
is  a self-adjointness
condition. The condition that the fibers $J^{-1}_i(x)$ be simply connected for
each $x\in P_i$
($i=1,2$) cannot be omitted, as will become clear from the proof of the next
theorem.

We recall that a symplectic realization of a Poisson manifold $P$ consists of a
symplectic
 manifold $S$
and a Poisson map $\rh:S\raw P$ \ci{Wei83,CDW}. This leads to a Poisson
morphism
$\rh^*:\cin(P)\raw\cin(S)$, which is the classical analogue of a representation
of a
$\mbox{}^*$-algebra on a Hilbert space \ci{NPLcq}.  There is an obvious
equivalence relation between
symplectic realizations, that is, $\rh_1:S_1\raw P$ and $\rh_2:S_2\raw P$ are
equivalent if there
exists a symplectic diffeomorphism $T: S_1\raw S_2$ such that $\rh_1=\rh_2\circ
T$.
In what follows, a realization will mean a symplectic one.
The following theorem  was proved by Xu in the special case (which covers most
cases of
physical interest) that the Poisson manifolds in question are integrable. His
proof (which is spread
out over sect.\ 4 of \ci{Xu89} and sect.\ 3 of \ci{Xu91}) follows the lines of
first showing that
integrable Morita equivalent Poisson manifolds have Morita equivalent
symplectic groupoids, which in
turn have equivalent categories of complete symplectic realizations. Our proof
below avoids the use
of symplectic groupoids, which may be a loss from a geometric point of view,
but has the advantage
of being similar in spirit to the proof of the imprimitivity theorem for
operator algebras.
\ci{Rie74,FD}.
\begin{theorem}
Let $P_1$ and $P_2$ be Morita equivalent Poisson manifolds. Then there is a
bijective correspondence
between their respective complete symplectic realizations. \ll{me}
\end{theorem}
{\em Proof.}   Given the equivalence bimodule  $P_2\stackrel{J_2}{\law}S
\stackrel{J_1}{\raw} P_1^-$,
there is a second  equivalence bimodule $P_1 \stackrel{J_1}{\law} S^-
\stackrel{J_2}{\raw} P_2$.
Given a realization $\rh:S_{\rh}\raw P_1$, one uses the former  equivalence
bimodule to obtain a
realization $J^{\rh}: S^{\rh}\raw P_2$, where $S^{\rh}=S*_{P_1}S_{\rh}/{\cal
F}$ is the symplectic
space constructed in Theorem \ref{cis}, and $J^{\rh}$ is given by
$J^{\rh}([x,y]_1)=J_2(x)$. Here
$[(x,y)]_1$ is the equivalence class of $(x,y)\in S\times S_{\rh}$ under the
foliation $\cal F$, and
the map $J^{\rh}$ is well-defined, because by the theory of full dual pairs
\ci{Wei83} the foliation
$\cal F$ restricted to $S$ coincides with the foliation by the fibers of $J_2$.
Also, the same fact
combined with the assumption that the quotient of $S$ by the $J_2$-foliation is
a manifold implies
that $S^{\rh}$ is a manifold.

We now relabel $S^{\rh}$ as $S_{\sg}$, and $J^{\rh}$ by $\sg$, and use the
second  equivalence
bimodule to find the corresponding induced  realization  $J^{\sg}: S^{\sg}\raw
P_1$.
Below we construct a symplectic diffeomorphism $V:S^{\sg}\raw S_{\rh}$, which
satisfies
$J^{\sg}=\rh\circ V$. Since all constructions evidently preserve completeness,
this establishes the
theorem.

Consider $(S *_{P_1} S_{\rh}) *_{P_2} S^- \subset S\times S_{\rh}\times S^-$,
that is, the space of
triples $(x,\th,y)$ satisfying $J_1(x)=\rh(\th)$ and $J_2(x)=J_2(y)$. The space
$S^{\sg}$ is obtained
from this by a double foliation: the first one $\F_1$ on $S\times S_{\rh}$
generated by the
Hamiltonian vector fields defined by the functions $J_1^*f-\rh^*f$,
$f\in\cin(P_1)$, and the second
one $\F_2$ on $S\times S^-$ generated by the Hamiltonian vector fields defined
by the functions
$\sg^*g-J_2^*g$, $g\in\cin(P_2)$. Let a triple $(x,\th,y)$ as above be given.
As above, we denote
equivalence classes defined by the first foliation by $[\cdot ,\cdot]_1$, and
those defined by the
second one by $[\cdot ,\cdot]_2$.

 We now
once again exploit the crucial fact from full  dual pairs that the foliation of
$S$ generated by the
the Hamiltonian vector fields defined by the functions $J_1^*f$,
$f\in\cin(P_1)$,  coincides with the
foliation by the fibers of $J_2$. Hence since $J_2(x)=J_2(y)$, we can find
$f\in\cin(P_1)$ for
which the flow $\phv_t$ of $X_{J^*f}$ satisfies $\phv_0(x)=x$, $\phv_1(x)=y$.
Let $\til{\phv}_t$ be
the flow of $-X_{J_{\rh}^*f}$ on $S_{\rh}$; by our assumption that $\rh$ be
complete, this flow
exists for all times, and we can define $\til{\th}=\til{\phv}_1(\th)$. By
standard foliation theory,
$\til{\th}$ only depends on $\th$ and the homotopy class in the fiber
$J_2^{-1}\circ J_2(x)$ of the
path $\{\phv_t\}_{t\in [0,1]}$ connecting $x$ and $y$. But this fiber is
assumed to be simply
connected, so that $\til{\th}$ is uniquely determined by $(x,\th,y)$.

We now define $V:S^{\sg}\raw S_{\rh}$ by $V([[x,\th]_1,y]_2)=\til{\th}$. This
is well-defined, and is
a symplectomorphism: given a triple $(x,\th,y)$ we have seen that we may choose
a representative
$(y,\til{\th},y)$ in the class $([x,\th]_1,y)$ defined by $\F_1$, and we
subsequently note that the
foliation $\F_2$ coincides with the foliation by the fibers of $J_1$. Since
$J_1(y)=\rh(\til{\th})$
is determined by $\til{\th}$, it follows that $V$ is a bijection. It is a
symplectomorphism by
Theorem \ref{cis}.

Finally, $ J^{\sg}([[x,\th]_1,y]_2)= J_1(y)= \rh(\til{\th})
= \rh\circ V
([[y,\til{\th}]_1,y]_2)=\rh\circ V ([[x,\th]_1,y]_2) $,
so that $J^{\sg}=\rh\circ V$, as announced. \Box

Note that  we could have weakened the
definition of a classical equivalence bimodule by omitting the manifold
condition on the foliations of
$S$ by $J_1$ and $J_2$ in Definition \ref{ceb}. In that case we would have
obtained a bijection
bewteen the set of realizations $S_{\rh}$ of $P_1$ for which $S^{\rh}$ is a
manifold, and the
analogous set defined for $P_2$.

Note, that \ci{KKS} and \ci{Wei83} already mention the fact that (in modern
parlance) the
symplectic leaves of Morita equivalent Poisson manifolds are in bijective
correspondence. This is
obviously a special case of Theorem \ref{me}, for the injection of a symplectic
leaf into its
Poisson manifold is of course a special instance of a symplectic realization
(in fact, such
realizations play a preferred role, in that they are irreducible in the sense
defined in
\ci{NPLcq}).
\subsection{Symplectic induction in stages}
After the imprimitivity theorem, the second most important and characteristic
result in Mackey's
theory of induced group representations is the theorem on induction in stages
\ci{Mac}.
This was generalized by Rieffel to his setting of induced representations of
$C^*$-algebras
\ci{Rie74,FD}. The symplectic counterpart is very easy, and the proof of the
following theorem
consists of simple bookkeeping, which we leave to the reader.

Let $J:S\raw P^-$ and $\rh:S_{\rh}\raw P$ be Poisson maps, with $S$ symplectic,
and let $\pi^{\rh}:A
\raw \cin(S^{\rh})$ be the corresponding induced representation of an
appropriate Poisson algebra
$A\subset J^*\cin(P^-)\subset \cin(S)$ (cf.\ Proposition \ref{cir}). Now assume
that the realization
$\rh$ is itself induced, in the sense that there are a Poisson manifold
$\til{P}$ and symplectic
manifolds $\til{S}$ and $S_{\sg}$, as well as Poisson  maps
$\til{J}:\til{S}\raw \til{P}^-$,
$\hat{J}:\til{S}\raw P$, and $\sg:S_{\sg}\raw \til{P}$, such that
$S_{\rh}\simeq \til{S}^{\sg}$ and
$\rh\simeq \til{J}^{\sg}$ (where $\til{J}^{\sg}: \til{S}^{\sg}\raw P$ is
constructed as in Theorem
\ref{cis} and the text following Proposition \ref{cir}, with
$S,P,P_2,J,J_2,S_{\rh},\rh$ replaced by
$\til{S},\til{P},P,\til{J},\hat{J},S_{\sg},\sg$, respectively).

Now form the symplectic manifold $S'=(S*_{P}\til{S})/\F$ as in Definition
\ref{def1}
(assuming that the leaf space of the foliation is indeed a manifold) and
Theorem \ref{cis} (that is,
$S*_P\til{S}$  consists of those pairs $(x,y)\in S\times \til{S}$ for which
$J(x)=\hat{J}(y)$, and
the foliation $\F$ is generated by $X_{J^*f}-X_{\hat{J}}^*f$, $f\in\cin(P)$).
\begin{theorem} With the above notation:\\
i) There is a well-defined
Poisson map $J':S'\raw \til{P}^-$ defined by $J'([x,y])=\til{J}(y)$, and a
Poisson morphism $\mu:
A\raw \cin(S')$ given by $(\mu(f))([x,y])=f(x)$.\\
ii) The induced symplectic space $(S')^{\sg}=(S'*_{\til{P}} S_{\sg})/\til{\F}$
constructed with
the maps $J'$ and $\sg$ is symplectomorphic to $S^{\rh}$, and the corresponding
induced
representation $(J')^{\sg}\circ\mu$ of $A$ on $\cin((S')^{\sg})$ is equivalent
to $\pi^{\rh}$ on
$\cin(S^{\rh})$.\\
iii) In the special case that one has a Poisson manifold $P_2$ and a Poisson
map $J_2: S\raw P_2$, so
that $A=J_2^*\cin(P_2)$, one has thus obtained a symplectic realization
$J^{\sg}: S'\raw P_2$ which
is equivalent to $J^{\rh}:S^{\rh}\raw P_2$.
\ll{tis}
\end{theorem}

It is worth spelling out the special case of Marsden-Weinstein reduction in
stages.

Take a
connected Lie group $G$ with closed connected subgroup $H\subset G$,
 and consider the actions $G\law T^*G\raw H$, being the pull-backs of the
action of $G$ on itself by
left-multiplication, and of the right-action of $H$ on $G$ by
right-multiplication. The
symplectic form on $T^*G$ is $\om=d\th_L$, with $\th_L$ the Liouville form.
This leads to two moment
maps $\g^*\stackrel{J_L}{\law} T^*G \stackrel{J_R}{\raw}(\h^*)^-$. Pick a
co-adjoint orbit $\O\subset
\h^*$, and form the reduced space $(T^*G)^{\O}=J_R^{-1}(\O)/H$. This produces a
Poisson map
$J_L^{\O}:(T^*G)^{\O}\raw \g^*$, which is just the moment map for the left
$G$-action on
$(T^*G)^{\O}$, which is inherited from the left $G$-action on $T^*G$.

In the left trivialization of
$T^*G\simeq G\times \g^*$   this reads as follows. The Liouville form is
$\th_L(x,p)=p_a\th^a(x)$
(where $\{\th_a\}_a$ is a basis of left-invariant one-forms on $G$),
$J_L(x,p)=xp$, $J_R(x,p)= p\upharpoonright \h $, with  $x\in G$ and
$p\in\g^*$; $xp$ denotes the
co-adjoint action of $x$ on $p$. Hence $(T^*G)^{\O}$ consists of equivalence
classes $[x,p]_H$, such
that $p\upharpoonright \h \in\O$; the equivalence relation is
$(xh,h^{-1}p)\sim (x,p)$ for all
$h\in H$. The induced $G$-action is $y[x,p]_H=[yx,p]_H$, and the moment map is
$J_L^{\O}([x,p]_H)=xp$.
All this can be found in \ci{MRW}.

Suppose $G$  acts on a symplectic manifold $S$ from the right in strongly
Hamiltonian
fashion, with associated moment map $J:S\raw (\g^*)^-$. We then induce from
$(T^*G)^{\O}$, obtaining
a symplectic space $S^{(T^*G)^{\O}}$, defined as usual: we start with
$S*_{\g^*}(T^*G)^{\O}=\{(s,z)\in S\times (T^*G)^{\O}|J(s)=J_L^{\O}(z)\}$, and
quotient by the
characteristic foliation, which in this case coincides with the foliation
generated by the $G$-action
$\rh$ given by  $\rh_x(s,z)=(sx,x^{-1}z)$. Hence
$S^{(T^*G)^{\O}}=(S*_{\g^*}(T^*G)^{\O})/G$.

On the other hand, we may restrict the $G$-action on $S$ to $H$, with moment
map $J_H:S\raw \h^*$
simply given by the restriction of $J$ to $\h$. This leads to the reduced space
$S^{\O}=J_H^{-1}(\O)/H$.
\begin{coro}
With the notations introduced above,  $S^{(T^*G)^{\O}}\simeq S^{\O}$. \ll{mwis}
\end{coro}
{\em Proof.} This follows from Theorem \ref{tis} above, with $P=\g^*$,
$\til{S}=T^*G$,
$\til{P}=\h^*$, $S_{\sg}=\O$, and $\sg=i_{\O}$ (the inclusion map of $\O$ into
$\h^*$). To obtain
Corollary \ref{mwis}, one only needs to verify that $(S*_{\g^*}T^*G)/G$ is
symplectomorphic to $S$,
which is Prop.\ A4 of \ci{Wei87}.

It may be instructive to give a direct proof, too. The induced space
$S^{(T^*G)^{\O}}$ consists of
equivalence classes $\mbox{}_G[s,x,p,\th]_H$, where the quadruple
$(s,x,p,\th)\in  S\times G\times
\g^*\times \O$ satisfies $J(s)=xp$ and  $p\upharpoonright \h =\th$. The
equivalence relation is
$(s,x,p,\th)\sim (ys,yxh^{-1},hp,h\th)$ for all $y\in G$ and $h\in H$.
It is then readily verified  that $W:S^{(T^*G)^{\O}}\raw J_H^{-1}(\O)/H$ given
by
$W(\mbox{}_G[s,x,p,\th]_H)=[x^{-1}s]_H$ defines a symplectomorphism. \Box

This corollary is not as academic as it may appear. As shown in subsections
4.1, 4.2, any
co-adjoint orbit of a nilpotent or linear semi-direct product Lie group $G$ is
of the form
$(T^*G)^{\O}$, so that Marsden-Weinstein reduced spaces with respect to such
groups can always be
obtained in a substantially simpler fashion by reducing with respect to an
appropriate subgroup $H$.
\section{Quantization of  the symplectic induction procedure}
\subsection{Rieffel induction}
\eo
The so-called Rieffel induction process, which we propose as the quantum
counterpart of
generalized Marsden-Weinstein reduction (``symplectic induction'') is discussed
in detail in
\ci{Rie74,FD}, so we will just recall the basic definitions and constructions.
Let $\A$ and $\B$ be $\mbox{}^*$-algebras which act on a Hilbert space $\H$
from the left and from
the right, respectively.
In physics, this situation corresponds
to having a quantization of the unconstrained system, as well as of the algebra
of constraints.
 $\B$ will always, and $\A$ will usually a  be pre-$C^*$-algebra or a
$C^*$-algebra, but it is possible (and necessary for some applications, cf.\
subsect. 4.5 below) to
take $\A$ to be an ${\rm Op}^*$-algebra of unbounded operators \ci{Sch}, that
is, a
 $\mbox{}^*$-algebra defined on a common dense domain $D\subset \H$; in that
case the space
$L\subseteq\H$ introduced below will have to lie in $D$.
The key ingredient of the induction process, playing the role of the
quantization of the moment map
in symplectic geometry, is a rigging map. This map, denoted by $\la \cdot,\cdot
\ra_{\B}$ is defined
on $L\times L$, where $L$ is a subspace of $\H$ (preferably dense, but this is
not strictly
necessary), which is mapped into itself under the action of $\A$ as well as
$\B$. The rigging map
takes values in $\B$, and must satisfy the following  conditions for all $\ps,
\phv\in L$:
\begin{enumerate}
\item $\la \lm\ps,\mu\phv\ra_{\B}=\ovl{\lm}\mu \la\ps,\phv\ra_{\B}$ for all
$\lm,\mu\in \Bbb C$;
\item $\la\ps,\phv\ra^*=\la\phv,\ps\ra_{\B}$;
\item $\la \ps,\phv B\ra_{\B}=\la\ps,\phv\ra_{\B}B$ for all $B\in\B$;
\item $ \la A\ps,\phv\rab=\la\ps,A^*\phv\rab$ for all $A\in\A$.
\end{enumerate}
Thus the rigging map is an operator-valued sesquilinear  product;   if it is
also positive in the
sense that $\la\ps,\ps\rab\geq 0$ for all $\ps\in L$, and if $L=\H$ with $\A$
and $\B$
$C^*$-algebras, then $\H$ equipped with the rigging map is called a Hilbert
$C^*$-module (for $\A$).

The aim of the Rieffel induction process is to obtain a representation
$\pug(\A)$ on some Hilbert
space $\hug$, given a representation $\plg(\B)$ on a Hilbert space $\hlg$. This
is possible if $\plg$
is $L$-positive in the sense that $\plg(\la\ps,\ps\rab)\geq 0$ for all $\ps\in
L$, as an operator on
$\hlg$. If so, one obtains $\pug$ in two steps: firstly, the algebraic tensor
product $L\ot\hlg$ is
formed, and endowed with a bilinear form $(\cdot ,\cdot)_0$, defined by
\be
(\ps\ot v,\phv\ot w)_0=(\plg(\la \phv,\ps\rab)v,w)_{\ch}, \ll{3.1}
\ee
where $(\cdot ,\cdot)_{\ch}$ is the inner product in $\hlg$ (taken linear in
the first entry, unlike
the rigging map; we follow the conventions of \ci{Rie74,FD}). This form is
positive semi-definite if
$\plg$ is $L$-positive. Secondly, one forms the quotient of $L\ot\hlg$ by the
subspace $\H_0\subset
L\ot\hlg$ of vectors with vanishing $(\cdot ,\cdot)_0$ norm, and completes the
quotient (equipped with
the form inherited from
 $(\cdot ,\cdot)_0$) into a Hilbert space $\hug$. (To make this procedure
resemble the formation of
the induced symplectic space $S^{\rh}$ in  Theorem \ref{cis} a little bit more,
one could follow
\ci{Rie74} in introducing the intermediate step of forming the tensor product
$L\ot_{\B}\hlg$, which
is the quotient of $L\ot\hlg$ by vectors of the type $\ps B\ot
v-\ps\ot\plg(B)v$, $B\in\B$, but by 3
and (\ref{3.1})  above such vectors are automatically of zero norm, so this
intermediate step is
incorporated in quotienting $L\ot\hlg$ by its null space. It is the latter step
which is obviously
the quantum analogue of the third step in forming $S^{\rh}$, namely the
quotienting by the null
foliation of the induced symplectic form on $S*_P S_{\rh}$).

Denoting the image of an elementary vector $\ps\ot v\in L\ot\hlg$ in the
completion $\hug$ of the
quotient $L\ot\hlg/\H_0$ by $\ps\til{\ot}v$, the representation $\pug(\A)$ is
then defined on the
subspace of $\hug$ of finite linear combinations of such images by
\be
\pug(A) \ps\til{\ot}v =(A\ps)\til{\ot}v, \ll{3.2}
\ee
compare with (\ref{2.a}). This representation is well-defined on account of  3
and 4 above.
If $\A$ is a (pre-) $C^*$-algebra, then the boundedness of $A\in\A$ does not
guarantee that
$\pug(A)$ is a bounded operator on $\hug$. On top of that,  it is necessary and
sufficient that the
bound
\be
\plg(\la A\ps,A\ps\rab) \leq\,  \n A \n^2 \plg(\la\ps,\ps\rab) \ll{3.2a}
\ee
 holds for all $\ps\in L$. A
stronger condition, implying this bound, is that the maps $T_{\ps}:\A\raw \B$
defined by $T_{\ps}(A)=
\la A\ps,\ps\rab$ are continuous for each $\ps\in L$.  This, in turn, is
implied if $\A$ and $\B$ are
$C^*$-algebras, and $T_{\ps}$ is positive (that is, $\la A\ps,A\ps\rab\geq 0$
in $\B$ for all
$\ps\in L$ and all $A\in\A$), for a positive map between $C^*$-algebras is
automatically continuous.
Of course, this would imply that the rigging map itself is positive (in the
sense explained after
the list of conditions above), so that any representation $\plg(\B)$ may be
used to induce from.  In
any case, if $\pug(A)$ is bounded for all $A$ in a pre-$C^*$-algebra $A$ one
may extend the induced
representation to the completion of $\A$. See \ci[Prop.\ 4.27]{Rie74},
\ci[XI.7.11-12]{FD},
\ci{Rie82}  for more information on these points.

It will be obvious to the reader that the (pre-) Hilbert space structure of $L$
has not been used at
all in this induction process, so that any linear space could have been used.
The reason we have
assumed that $L\subset\H$ is a (dense) subspace of a Hilbert space is that this
is the setting in
which the Rieffel induction procedure will be used in the quantization of
constrained systems, and
the main difficulty is then to identify $L$ and the rigging map, given $\H$ and
the actions of $\A$
and $\B$ on it.

The form of $\hug$ as given is useful for the computation of physical
correlation functions (that
is, expectation values of (time-ordered) products of the type
$(\pug(A_1(t_1))\ldots
\pug(A_n(t_n))\Om,\Om)$, where $A_i\in\A$, and $\Om$ is some physically
relevant state in $\hug$),
which can be evaluated in $L\ot\hlg$ on  any pre-image of $\Om$, using the
inner product $(\cdot\,
\cdot)_0$; the contributions of intermediate states with zero norm will
automatically drop out.
Nonetheless, it is useful to have an alternative realization of $\hug$
\ci{Han}. Let $\ovl{L}$ be the
conjugate space of $L$ (which coincides with $L$ as an additive group, but has
the conjugate scalar
multiplication), and let ${\cal L}(\ovl{L},\hlg)$ be the space of linear maps
of $\ovl{L}$ into
$\hlg$. Then define $U:L\ot\hlg\raw {\cal L}(\ovl{L},\hlg)$ by
\be
(U(\ps\ot v))(\phv)=\plg(\la\phv,\ps\rab)v. \ll{3.3}
\ee
One can define an inner product $(\cdot,\cdot)^{\ch}$  on the image $\Im$ of
$L\ot\hlg$ in ${\cal
L}(\ovl{L},\hlg)$ under $U$ by
\be
(U(\ps\ot v), U(\phv\ot w))^{\ch}=( \plg(\la\phv,\ps\rab)v,w)_{\ch}; \ll{3.4}
\ee
this form is positive definite, and the closure of $\Im$ in this inner product
yields a Hilbert space
$\til{\H}^{\ch}$. Noticing that $U$ exactly annihilates $\H_0\subset L\ot\hlg$,
it follows that $U$
quotients and extends to a well-defined unitary operator
$\til{U}:\hug\raw\til{\H}^{\ch}$.

We continue by recalling Rieffel's generalized imprimitivity theorem
\ci{Rie74,FD,Rie82}, which we will actually use later on, and whose explicit
form will make it clear
that the symplectic  imprimitivity theorem (Theorem 2 in subsect.\ 2.2) is
indeed a `classical'
version of the former. We assume that $\A$ and $\B$ are (pre-) $C^*$-algebras,
acting on $L$ as
above, which is equipped with a rigging map $\la \cdot,\cdot\rab$ satsfying all
properties stated
earlier.  $L$ is called an $\A - \B$ imprimitivity bimodule (at least in
\ci{Rie74,FD}; later the
terminology `equivalence bimodule' was adopted \ci{Rie82}) if in addition there
is  a rigging map
$\mbox{}_{\A}\la\cdot,\cdot\ra:L\times
L \raw \A$, satisfying the same properties of the $\B$-rigging, but with the
roles of $\A$ and $\B$,
and left and right interchanged. Moreover, the following conditions must
hold:\\
i) the bounds (\ref{3.2a}), as well as the corresponding ones with $\A$ and
$\B$ interchanged, hold;
\\ ii) the linear span of $\{\la \ps,\phv\rab | \ps,\phv\in L\} $ is dense in
$\B$, and similarly with
$\B$ replaced by $\A$;\\
iii) $\mbox{}_{\A}\la \ps,\phv\ra\zt = \ps\la\phv,\zt\rab$ for all
$\phv,\ps,\zt\in L$.

The imprimitivity theorem   states that if there exists an $\A\, -\,  \B$
imprimitivity bimodule
(in which case $\A$ and $\B$ are called strongly Morita equivalent) then there
is a bijective
correspondence between the set of $L$-positive representations of $\A$ and $\B$
(which bijection
preserves a number of properties of representations, such as direct integrals
and weak containment,
but upsets others, such as cyclicity \ci{Rie74,FD}). The representation of $\A$
associated with
$\plg(\B)$ is simply $\pug$, given by the Rieffel induction process. To go in
the opposite direction,
one makes the conjugate space $\ovl{L}$ into a right- $\A$-module and left-
$\B$-module by
conjugating the respective actions on $L$, and induces using $\ovl{L}$ and the
$\A$-rigging map
$\mbox{}_{\A}\la \cdot,\cdot\ra$. This conjugation is analogous to the step in
the proof of the
symplectic imprimitivity theorem where one passes from $S$ to $S^-$.

 More generally, there is a striking formal correspondence between
(quantum) imprimitivity bimodules and classical equivalence bimodules (cf.\
Definition \ref{ceb}).
As already mentioned, the rigging map corresponds to the moment map, and the
compatibility condition
iii) (which implies that the actions of $\A$ and $\B$ on $L$ commute
\ci[XI.6.2]{FD}) replaces the
symplectic assumption that $J_1^*\cin(P_1)$ Poisson commutes with
$J_2^*\cin(P_2)$. Assumption ii)
is the quantum analogue of the part of the definition of a full dual pair which
states that $J_1$
and $J_2$ are surjective. The symplectic assumption that the leaf spaces of the
foliations defined by the
fibers of $J_1$ and $J_2$ are manifolds has its analogue in a condition which
we omitted in order to
state the imprimitivity theorem in its fullest generality; we could add\\
iv) the $\A$- and $\B$- rigging maps are positive\\
(that is, $\la\ps,\ps\rab\geq 0$ in $\B$ and  $\mbox{}_{\A}\la \ps,\ps\ra\geq
0$ in $\A$ for all
$\ps\in L$); if this condition is added the imprimitivity theorem  evidently
states that there is a
bijective correspondence between all representations of $\A$ and $\B$.
Conversely, the imprimitivity
theorem following from i)-iii) alone is analogous to the weakened version of
Theorem \ref{me} stated
following its proof.

If only a right $\B$ -module $L$ is given, together with a positive rigging map
whose image in dense
in $\B$, one can always find a  $C^*$-algebra $\A$ acting on $L$ so that $L$
becomes an $\A - \B$
imprimitivity bimodule \ci{Rie74,FD}.  This algebra $\A$ (called the
imprimitivity algebra of
$(L,\B)$) is generated by operators of the form $T_{(\ps,\phv)}$, whose action
on $\zt\in L$ is
defined by $T_{(\ps,\phv)}\zt=\ps \la\phv,\zt\rab$. Similarly, given a  Poisson
manifold $ P_1$ which
is one half of a full dual pair with equivalence bimodule $S$, one can find the
manifold $P_2$
completing the dual pair by taking the  Poisson commutant of $J^*\cin(P_1)$ in
$\cin(S)$, which is
necessarily of the form $J_2^*\cin(P_2)$, at least in the finite-dimensional
case. However, the
imprimitivity algebra $\A$ only coincides with the commutant of $\B$ if $L$ is
finite-dimensional (in
general it is not even a von Neumann algebra). This dichotomy between the
classical and the quantum
settings will presumably disappear if one studies infinite-dimensional Poisson
manifolds and their
Morita equivalence. \subsection{Quantum Marsden-Weinstein reduction}
We first apply the above framework to the quantization of the  symplectic
reduction procedure in its
original version, where one reduces by a group action (cf.\ the Introduction,
and the par.\ following
Prop.\ \ref{cir}). Hence  we assume that the classical data consisting  of a
symplectic manifold $S$,
a strongly Hamiltonian (right) action of a Lie group $H$ on $S$,  a Poisson
algebra $A\subset\cin(S)$
of functions which are invariant under the group action, and a co-adjoint orbit
$\O\subset\h^*$, have
been quantized as a Hilbert space $\H$, a unitary  representation $\pi(H)$ on
$\H$, a
representation of a $C^*$-algebra $\A$ on $\H$, which commutes with $\pi$, and
an  irreducible
unitary representation $\plg(H)$ on a Hilbert space $\hlg$, respectively.  (At
no cost one may
replace the co-adjoint orbit $\O$ and the irreducible representation $\plg$
 by an arbitrary symplectic space with a strongly Hamiltonian $H$-action and an
arbitrary
unitary representation of $H$, respectively. Moreover, in what follows $H$ does
not need to be a Lie
group; local compactness suffices.) Of course, the right $H$-action on $S$
amounts to a Poisson
morphism $J^*:\cin(\h^*)^-\raw\cin(S)$, and the representation $\pi(H)$ on $\H$
corresponds to an
anti-representation (called $\pi^-$) of the group algebra $C^*(H)$ \ci{Ped},
defined by
\be
\pi^-(f)=\int_H dh\, f(h)\pi(h^{-1}), \ll{3.5}
\ee
where $dh$ is the Haar measure on $H$ (assumed unimodular for notational
simplicity), and $f\in
C_c(H)$. Thus the $C^*$-algebra
 $C^*(H)$, being the appropriate completion of the convolution algebra $C_c(H)$
(playing the role of  $\B$ of the preceding subsection), is to be seen as the
quantization of the Poisson algebra $\cin(\h^*)$, a point of view first stated
by Rieffel
\ci{Rie89a}.
We remark that it has been proved  that $C^*(H)$ is a strict deformation
quantization of $\cin(\h^*)$
for $H$ nilpotent \ci{Rie89b} or compact \ci{NPLstr}, and we expect it to be
true for any amenable
group.

Let us first assume that $H$ is compact, with Haar measure normalized to unity.
We then take $L=\H$,
$\B=C_c(H)$, and define the rigging map by
\be
\la \ps,\phv\ra_{C_c(H)}:\, h\raw (\pi(h)\phv,\ps), \ll{3.6}
\ee
utilizing the inner product in $\H$. This is easily shown to satisfy all
conditions stated in the
previous subsection, and it is positive as well:
\begin{lemma}  For $H$ compact, $\la \ps,\ps\ra_{C_c(H)}\geq 0$ as an
element of $C^*(H)$ for all $\ps\in\H$.\ll{pos}
\end{lemma}
{\em Proof}. Let $\bone_H$ denote the function on $H$ which is
identically equal to one. Then $\bone_H^* *\bone_H=\bone_H$ (where $*$ is the
convolution product on
$C(H)$), so that $\bone_H$ is a positive element of $C^*(H)$. Hence for any
representation
$\til{\pi}(H)$ on $\til{\H}$ (with inner product $(\cdot,\cdot)_{\sim}$)
$\int_H dh\,
(\til{\pi}(h)\til{\ps},\til{\ps})_{\sim}=(\til{\pi}(\bone_H)\ps,\ps)_{\sim}\geq
0$ for all
$\til{\ps}\in\til{\H}$. Now choose $\pi_1$ an arbitrary unitary representation
of $H$ on $\H_1$
(with inner product $(\cdot,\cdot)_1$). Using the previous argument with
$\til{\pi}=\pi\ot\pi_1$ and
$\til{\ps}=\ps\ot\ps_1$, we find that $(\pi_1(\la
\ps,\ps\ra_{C_c(H)})\ps_1,\ps_1)_1\geq 0$ for all
$\ps_1\in\H_1$. Since $\pi_1$ was arbitrary, this proves the lemma.  \Box

Therefore, any unitary representation $\plg$ of $H$ may be used to induce from.
This is remarkable,
for it implies that for compact Lie groups there is no quantum analogue of
singular values of the
moment map.
Moreover,  any $C^*$-algebra contained in the commutant $\pi(H)'$ of $\pi(H)$
on $\H$
is represented by bounded operators in the representation $\pug$ on the Hilbert
space $\hug$.
This follows from
\begin{lemma}
If $A\in \pi(H)'$ and $H$ compact then $\la A\ps,A\ps\ra_{C_c(H)} \leq \n\,
A\n^2\, \la\ps,\ps
\ra_{C_c(H)}$ in $C^*(H)$ for all $\ps\in\H$. \ll{bound}
\end{lemma}
{\em Proof.} Notation as in the proof of the previous lemma.  That proof showed
that the operator
$P=\int_H dh\, \pi\ot\pi_1(h)$ is positive on $\H\ot\H_1$.  Clearly,
$P$ commutes with $A\ot \Bbb I$ if
$A\in \pi(H)'$. Hence with $\til{\ps}=\ps\ot\ps_1$ and $(\cdot,\cdot)_{\ot}$
the inner product in
$\H\ot\H_1$,   \be
(P A\ot \Bbb I \til{\ps}, A\ot \Bbb I\til{\ps})_{\ot}\leq \n A\n^2\, \n
P^{1/2}\til{\ps}\n^2=
\n A\n^2\, (P\til{\ps},\til{\ps})_{\ot}. \ll{3.in}
\ee
If $\om$ denotes the state on $C^*(H)$ defined by $\ps_1$, then this inequality
reads\\
$\om(\la A\ps,A\ps\ra_{C_c(H)})\leq \n A\n^2\, \om(\la \ps,\ps\ra_{C_c(H)})$,
which proves the
lemma. \Box

Let us see what  the trivially induced representation looks like. We take
$\hlg=\Bbb C$,carrying the
trivial representation of $H$, so that the space $L\ot\hlg$ used in the
construction is simply $\H$.
Using (\ref{3.1}) and (\ref{3.6}), we find that $(\ps,\ps)_0=(P_{\rm
id}\ps,P_{\rm id}\ps)$, where
$P_{\rm id}$ is the orthogonal projector on the subspace $\H_{\rm id}\subset\H$
(which may be empty)
of vectors which are invariant under $H$. The null space $\H_0$ is the
orthogonal complement of
$\H_{\rm id}$, and the final induced space $\H^{\rm id}=\H/\H_0$ is simply
$\H_{\rm id}$, with the
original inner product of $\H$. This space is invariant under $\pi(H)'$, so we
find that $\pug(A)$ on
$\H_{\rm id}$ is just the restriction of $A$ to  $\H_{\rm id}$. This is, of
course, nothing but
Dirac's prescription \ci{Dir} for first-class constraints (it goes without
saying that the above
procedure quantizes the Marsden-Weinstein reduced space at zero, so that all
the classical
constraints are indeed first-class).

The Dirac procedure breaks down if zero is not in the discrete spectrum of each
of the constraints, a
situation which may arise when $H$ is non-compact. The Rieffel induction
procedure can still be used
in that case, the main problem being the identification of an appropriate
subspace $L\subset \H$.
This will have to be done case by case; for example, if $M$ is a manifold on
which $H$ acts, and
$\H=L^2(M)$, one may take $L=C_c(M)$. With $\B=C_c(H)$ and the rigging map
still given by
(\ref{3.6}), the conditions on $L$ are simply that the rigging map indeed takes
values in $C_c(H)$,
and that $L$ is mapped into itself by $\A$ and $\B$. (In favourable
circumstances, one may be able to
extend these mappings by continuity to $L=\H$, and the respective
$C^*$-closures of $\A$ and $\B$.)

 In the following proposition, the assumption of unimodularity is only made for
convenience (in the general case the rigging map and the convolution product
would contain the modular
function of $H$).
\begin{prop}
Let $H$ be locally compact and unimodular, and let $L$ be such that
(\ref{3.6}) defines a function in
$C_c(H)$ for all $\ps,\phv\in L$. Then $\la\cdot,\cdot\ra_{C_c(H)}$ is a
rigging map, which is
positive if $H$ is amenable. Whether or not $H$ is  amenable, every
representation of $H$ weakly
contained in the regular one is $L$-positive (so that it may be used to induce
from). \ll{rmp}
 \end{prop}
{\em Proof}. The verification of properties 1-4 of a rigging map (cf.\ previous
subsect.) is
trivial. As to the positivity, the proof of Lemma \ref{pos} clearly breaks down
in the noncompact
case, as the function $\bone_H$ is not in $C_c(H)$ (or, indeed, in $C^*(H)$).
However, if $H$ is
amenable it has a family of subsets called $\{U_j\}_{j\in J}$ in \ci[3.6]{Gre}
(where our $H$ is
called $G$).  Here $J$ is a directed index set, and the $U_j$ eventually fill
up $H$. Each $U_j$ is
measurable and has   finite Haar measure $\mu(U_j)$, and one has the following
property. We define a
family of functions $g_j\in L^1(H)\subset C^*(H)$ by
$g_j=(\mu(U_j))^{-1/2}\ch_{U_j}$ (with $\ch_E$
the characteristic function of a Borel set $E$). Then $\lim_j g_j *
g_j^*=\bone_H$ pointwise on $H$.
Hence for any $f\in L^1(H)$ one has by the bound $g_j * g_j^*\leq \bone_H$ and
the Lebesgue dominated
convergence theorem that $\lim_j \int_H dh\, f(h) g_j * g_j^*(h)=\int_H dh\,
f(h)$. Clearly, each
$g_j * g_j^*$ is a positive element of $C^*(H)$. (These results  easily follow
from \ci[3.6]{Gre}, and
are even given as the definition of amenability in \ci[II.3]{Ren}, specializing
the groupoids in
this ref.\ to groups.) Using the notation and strategy of the proof of Lemma
\ref{pos}, we now take
$f(h)=(\pi\ot\pi_1(h)\til{\ps},\til{\ps})_{\sim}$. Then  $$\int_H dh\, f(h) g_j
* g_j^*(h)=
(\pi\ot\pi_1(g_j * g_j^*)\til{\ps}, \til{\ps})_{\sim}\geq 0$$ for all $j$.
Therefore, $\int_H dh\,
f(h)\geq 0$. As in the proof of
 Lemma \ref{pos}, we   conclude that $\la \ps,\ps\ra_{C_c(H)}\geq 0$ in
$C^*(H)$.

If $H$ is not amenable, the family  $\{U_j\}_{j\in J}$ with the desired
properties does not exist.
However, in that case $\la \ps,\ps\ra_{C_c(H)}$ is a positive element of the
reduced group algebra
$C_r^*(H)$ by an argument due to Rieffel \ci{Rie88} (in particular his
calculation (1.1),
specialized to $A=\Bbb C$). The proposition follows. \Box
\begin{prop}
In Lemma \ref{bound} above one may replace `compact' by `amenable'.\ll{replace}
\end{prop}
{\em Proof}. This can be proved in a similar way as Lemma \ref{bound},
replacing the operator $P$
by $P_j=\int_H dh\, \pi\ot\pi_1(h) g_j * g_j^*(h)$, which is well-defined since
$ g_j * g_j^*$ has
compact support.  One then obtains (\ref{3.in}) with $P$ replaced by $P_j$, and
taking the limit in
$j$ yields the proposition. \Box

Thus one may encounter a quantum analogue of a singular value of the moment map
if $H$ is, for
example, non-compact and semi-simple, and the representation $\plg$ one induces
from does not
contribute to the Plancherel measure. Assuming, instead, that we are in the
regular case (that is,
$\plg(H)$ is $L$-positive), we are now in a position to illustrate (\ref{3.3})
and (\ref{3.4}).
Namely, let a Lie group $H$ act on a manifold $M$; the pull-back action on
$T^*M$ is then
automatically strongly Hamiltonian \ci{AM} with moment map $J$. For any
realization $\rh:
S_{\rh}\raw \h^*$ one may define the induced space $(T^*M)^{\rh}$ constructed
in subsect.\ 2.1. In the
special case where $\O$ is a co-adjoint orbit in $\h^*$ we thus obtain the
Marsden-Weinstein reduced
space $J^{-1}(\O)/H$ (assuming regularity).

The quantization of this setting is to take $\H=L^2(M)$ (the Mackey Hilbert
space of a manifold
\ci{AM}), carrying the obvious unitary representation $\pi(H)$ derived from the
(right) action of
$H$ on $M$. For simplicity, we assume that $M$ has an $H$-invariant measure
$\nu$ (if not, one works
with half-densities on $M$), so that $\H=L^2(M,\nu)$.  Choosing $L=C_c(M)$, the
rigging map
(\ref{3.6}) is then simply given by
\be
\la\ps,\phv\ra_{C_c(H)}: h\raw \int_M d\nu(m)\, \phv(mh)\ovl{\ps(m)}. \ll{3.8}
\ee
We now pick an $L$-positive representation $\plg$ of $H$, defined on a Hilbert
space $\hlg$
(as we saw above, for $H$ amenable any representation will do).
If $\ps\ot v\in C_c(M)\ot\hlg$ then the image $U(\ps\ot v)$ in ${\cal
L}(\ovl{L},\hlg)$ may be
identified with the $\hlg$-valued function $\ps_v$ on $M$ defined by
\be
\ps_v(m)=\int_H dh\, \ps(mh)\plg(h)v \ll{3.9}
\ee
(cf.\ Thm.\ 5.12 in \ci{Rie74} for the special case $M=G$ a group with
$H\subset G$ a subgroup).
This function satisfies the equivariance condition
$\ps_v(mh)=\plg(h^{-1})\ps(m)$ for all $m\in M$
and $h\in H$, and the inner product in $\til{\hug}$ of two such functions is
given by (\ref{3.4}).
This may be rewritten in terms of a so-called approximate cross-section of
$M/H$ in $M$, that is, a
continuous positive function $b$ on $M$ whose support $S$ is such that $S\cap
KH$ is compact
for any compact $K\subset H$ (here $KH=\{Kh|h\in H\}$), and $\int_H dh\,
b(mh)=1$ for all $m\in
M$.
Such a function is shown to exist in Lemma 1.2 of \ci{Mos}; for $M=G$, $b$ is
the Bruhat
approximate cross-section used in Theorem 4.4 in \ci{Rie74}. A short
computation then shows that the
inner product in   $\til{\hug}$ is given by
\be
(\ps_v,\phv_w)=\int_M d\nu(m)b(m)(\ps_v(m),\phv_w(m))_{\ch}, \ll{3.10}
\ee
where, as before, $(\cdot,\cdot)_{\ch}$ is the inner product in $\hlg$.
Alternatively, this may be
written as an integral over $M/H$ in terms of a suitable measure on that space,
for
 $(\ps_v(m),\phv_w(m))_{\ch}=(\ps_v(mh),\phv_w(mh))_{\ch}$ on account of the
equivariance condition
stated above. This leads to the generalized induced representations of
Moscovici \ci{Mos} (which
were already mentioned in \ci{Rie74} as a special case of the Rieffel induction
process).
In conclusion, the space  $\til{\hug}$ consists of $H$-equivarant functions
$\Psi$ on $M$ with values
in $\hlg$, such that $m\raw (\Psi(m),v)_{\ch}$ is measurable for each
$v\in\hlg$, and $(\Psi,\Psi)$
defined by (\ref{3.10}) is finite. Operators $A$ on $L^2(M)$ commuting with
$\pi(H)$ are then
naturally defined on  $\til{\hug}$ also, that is, the desired
induced representation is defined by $\pug(A)\ps_v=(A\ps)_v$. Hence we have
shown how the Moscovici construction follows from (\ref{3.3}) and
(\ref{3.4}), and it has been made clear of which symplectic situation it is the
quantization.

If we take $M=G$ and $H$ a closed   subgroup of $G$, acting on the latter from
the right, we find that
the rigging map (\ref{3.6}), defined on $L=C_c(G)$, is just the convolution
(over $G$) $\ps *\phv$,
restricted to $H$. The right-action  (\ref{3.5}) is just $\pi^-(f)\ps=\ps*f$
(convolution over $H$).
Hence this rigging map and right-action, which were directly defined by Rieffel
\ci{Rie74} in the
form just given, are specializations  of the general formulae (\ref{3.5}),
(\ref{3.6}). As detailed in
\ci{Rie74,FD}, the Rieffel induction procedure applied to this special case is
equivalent to Mackey's
formalism of induced group representations \ci{Mac,Var}.
Note, that  in this case the rigging map is positive even if $H$ is not
amenable (a fact \ci{Rie74}
 not covered by our Proposition \ref{rmp}).
 \subsection{Quantization of symplectic group actions which are not strongly
Hamiltonian}
 What happens
when the moment map $J:S\raw (\h^*)^-$ is not equivariant with respect to the
co-adjoint
representation $\pco$? (In the literature, one finds the notation ${\rm
Ad}^*_{h^{-1}}$ for our
$\pco(h)$.) Equivalently, the pull-back $J^*:\cin(\h^*)^-\raw \cin(S)$ fails to
be a Poisson morphism
with respect to the Lie-Poisson structure on $\h^*$ in that case. It is well
known how to handle this
situation in the classical case \ci{AM,GS}. The Lie group $H$, assumed to act
on $S$ from the right,
preserving the symplectic form and admitting a moment map, also acts on
$\cin(S)\ot \h^*$ by a
left-action $\al$ defined on $f\in\cin(S)\ot \h^*$ as follows: $(\al_h
f)(s)=\pco(h)f(sh)$. The
infinitesimal action $d\al$ of $X\in\h$ is then  $d\al_X
f=(\til{X}+d\pco(X))f$, where $\til{X}$ is
the vector field on $S$ defined by $(\til{X}f)(s)=d/dt\, f(s\exp(tX))_{|t=0}$.
Subsequently, define
an element $\Sigma\in\h^*\ot\h^*$ by $ \Sigma (X, Y)=\la (d\al_X J)(s),Y\ra$,
which is
independent of $s\in S$ (assuming $S$ connected).  Moreover, $\Sigma$ turns out
to be antisymmetric,
and defines a 2-cocycle on $\h$. Hence one may define a new Poisson bracket
$\{\cdot,\cdot
\}^{\Sigma}$ on $\cin(\h^*)$ by putting \be
\{\til{X},\til{Y}\}^{\Sigma}=\widetilde{[X,Y]}+\Sigma(X,Y)\bone_{\h^*};
\ll{3.11} \ee here
$\til{X}\in\cin(\h^*)$ is defined by $\til{X}(\th)=\la \th,X\ra$ (giving the
Poisson bracket on such
functions determines it completely), and $\bone_{\h^*}$ is the function which
is identically 1 on
$\h^*$. Then $J$ is a Poisson map with respect to this modified Poisson
structure of $\h^*$,  and in
addition is equivariant relative to the originally given $H$-action on $S$, and
the new $H$-action
$\pco^{\Sigma}$ on $\h^*$ defined by   \be
\pco^{\Sigma}(h)\th=\pco(h)(\th+J(s))-J(sh^{-1}),\ll{3.12}
\ee
 which
is independent of $s$.
Clearly, if $J$ was $\pco$-equivariant (that is, $\al_h J=J$ for all $h\in H$)
then $\Sigma=0$, and
(\ref{3.11}) reduces to the Lie-Poisson bracket.

The essential point is that the Poisson structure on $\cin(\h^*)$, originally
defined by the Lie
bracket on $\h$, is modified by a certain central extension $\Sigma$ of $\h$;
the moment map remains
the same. Also, the Marsden-Weinstein reduction with respect to a point
$\mu\in\h^*$ of $S$ is
practically unmodified (cf.\ exercise 2.4.3D in\ci{AM}), and is a special case
of the general
procedure described in subsect.\ 2.1, taking $S_{\rh}$ to be the symplectic
leaf of $\h^*$
containing $\mu$ (relative to the $\Sigma$-Poisson bracket), or equivalently,
the orbit of $\mu$
under the $H$-action (\ref{3.12}).

 This remark suggests how the situation should be quantized. Firstly, the
quantum analogue of a
symplectic group action which is not strongly Hamiltonian is a projective
unitary representation on
a Hilbert space $\H$, for by Wigner's theorem \ci{Var} that is the most general
structure which
quotients to a group action on the state space of $\H$ (i.e., the corresponding
projective space),
preserving the symplectic structure of the latter (defined by the inner product
on $\H$
\ci{AM,TW,NPLcq}). Thus we assume that for each $h\in H$ we are given a unitary
operator $\pi(h)$ on
$\H$, such that $\pi(h_1h_2)=c(h_1,h_2)\pi(h_1)\pi(h_2)$, where
$|c(h_1,h_2)|=1$ and the identity
$c(h_1,h_1)c(h_1h_2,h_3)=c(h_1,h_2h_3)c(h_2,h_3)$ is satisfied (this is the
complex conjugate of
the equation one obtains by demanding associativity of the $\pi(h)$). We say
that $\pi$ has
multiplier $\ovl{c}$ \ci{Var}. If this is seen as the quantization of the
$H$-action on $S$, one
expects that the infinitesimal version of $c$, that is, the 2-cocycle on $h$
derived from it,
coincides with $\Sigma$. Conversely, starting from $\Sigma$ one may attempt to
find a 2-cocycle $c$
on $H$ satisfying this property, which is always possible if $H$ is simply
connected (in general, a
certain quantization condition must be satisfied by $\Sigma$ \ci{TW}).

The quantum analogue of $\cin(\h^*)$ equipped with the Poisson bracket
(\ref{3.12}) is the twisted
group algebra $C^*(H,c)$ of $H$, which has a product (defined on $C_c(H)$ to
start)
\be
(f*_cg)(h)=\int_H dk\, f(hk^{-1})g(k)c(hk^{-1},k), \ll{3.13}
\ee
and involution
\be
(f^{*_c})(h)=c(h,h^{-1})\ovl{f(h^{-1})}. \ll{3.14}
\ee
We obtain a right-representation $\pi^-$  of $C^*(H,c)$ by
\be
\pi^-(f)=\int_H dh\, f(h)c(h,h^{-1})\pi(h^{-1}). \ll{3.15}
 \ee
There is a subtle difference with the untwisted case: there one can find both a
representation of $C^*(H)$ on $\H$ (obtained by replacing $h^{-1}$ in
(\ref{3.5}) by $h$), and a
right-representation, given by (\ref{3.5}). In the twisted case, one obtains a
representation of
$C^*(H,\ovl{c})$ by omitting $c$ and changing $h^{-1}$ to $h$ in (\ref{3.15}).
This is just as well,
as we will see in Proposition \ref{twpos} below.

Taking $\B=C_c(H,c)$ (that is, the restriction of $C^*(H,c)$ to its subspace
$C_c(H)$) in the
Rieffel induction process, we can define the rigging map by (\ref{3.6}), as in
the untwisted case,
and (repeatedly using the cocycle identity on $c$) easily check it satisfies
all conditions (assuming
that an appropriate subspace $L$ can be found). Moreover:
\begin{prop}
Let the locally comapct group $H$ be  amenable. Then the rigging map
(\ref{3.6}) is
positive (i.e., $\la\ps,\ps\ra_{C_c(H,c)}\geq 0$
 in $C^*(H,c)$). \ll{twpos}
\end{prop}
{\em Proof.} Also in the twisted case  there exists a one-to-one correspondence
between
representations $\pi_1$ of $C^*(H,c)$ on a Hilbert space $\H_1$ and projective
unitary
representations (called $\pi_1$ as well) of $H$ with multiplier $c$ \ci{Green};
the correspondence is
$\pi_1(f)=\int_H dh\, f(h)\pi_1(h)$, as in the untwisted case. Hence it is
sufficient to prove that
$(\pi_1(\la\ps,\ps\ra_{C_c(H,c)})\ps_1,\ps_1)_1\geq 0$ for all $\ps\in L$ and
$\ps_1\in\H_1$.
As we remarked above, $\pi$ is a representation of $H$ with multiplier
$\ovl{c}$, whereas $\pi_1$
has multiplier $c$. Hence $\pi\ot\pi_1$ is a representation of $H$ and
$C^*(H)$, without any
multiplier. Therefore, the argument used to prove Lemma \ref{pos} Proposition
\ref{rmp} applies.
Taking the compact case for simplicity, we can write
$$(\pi_1(\la\ps,\ps\ra_{C_c(H,c)})\ps_1,\ps_1)_1=(\pi\ot\pi_1(\bone_{\h^*})\ps\ot\ps_1,\ps\ot\ps_1)\geq
0,$$
now regarding $\ps\ot\ps_1$ as a representation of $C^*(H)$, in which
$\bone_{\h^*}$ is a positive
element. The noncompact case is handled exactly as in the proof of Proposition
\ref{rmp}. \Box

It is interesting to exhibit Rieffel's treatment of induced projective
representations \ci{Rie74} as
(almost) a special case of the above (cf.\ the discussion closing the previous
subsection). Namely,
assume that $H\subset G$ is a closed subgroup, with a multiplier $c$ given,
whose restriction to
$H$ is what we called $c$ before. Now take $\H=L^2(G)$ with $L=C_c(G)$ as the
dense subspace on which
the rigging map is defined. Then $(\pi(h)\ps)(x)=\ovl{c(x,h)}\ps(xh)$ defines a
projective unitary
representation of $H$ on $\H$ with multiplier $\ovl{c}$. Then (\ref{3.15})
specializes to
$\pi^-(f)\ps=\ps*_c f$ (convolution over $H$), whereas the rigging map
(\ref{3.6}) becomes
$\la\ps,\phv\ra_{C_c(H,c)}=\ps^{*_{\ovl{c}}}*_c \phv$ (convolution over $G$;
here
$\ps^{*_{\ovl{c}}}$ is defined by changing $c$ to $\ovl{c}$ in (\ref{3.14})).
By associativity of
$*_c$, the condition $\la\ps,\pi^-(f)\phv\ra_{C_c(H,c)}
=\la\ps,\phv\ra_{C_c(H,c)})*_c  f$ is
manifestly satisfied. Rieffel's right action of $C^*(H,c)$ is the one given
above, while his rigging
map is obtained by putting  $\la\ps,\phv\ra_{C_c(H,c)}=\ps^{*_{c}}*_c \phv$,
which is  positive even
if $H$ is not amenable (although not manifestly so, despite appearances, for
the convolution product
is in $C^*(G,c)$ rather than $C^*(H,c)$). We conjecture that the rigging map
coming from our approach
is positive in general as well.
 \subsection{Induction with groupoid algebras}
So far, the general formalism to quantize constrained systems has only been
illustrated for the case
that the Poisson algebra of the constraints is essentially a Lie algebra,
perhaps with central
extension. In other words, we took the Poisson algebra $\cin(P)$ to be
$\cin(\h^*)$, where $\h$ is a
Lie algebra; the quantization involved the group algebra $C^*(H)$ (perhaps
twisted).
A much more general situation that we are able to handle, in the sense that an
explicit formula for
the rigging map can be given, arises when we merely assume that $\h$ is a Lie
algebroid $\L(\Gm)$ of
a Lie groupoid $\Gm$ \ci{CDW}, and $\cin(P)\equiv \cin(\L(\Gm)^*)$ the Poisson
algebra canonically
associated to $\L(\Gm)$ \ci{CDW}. For we know \ci{NPLstr} that the quantization
of the Poisson
algebra $\cin (\L(\Gm)^*)$ is the groupoid $C^*$-algebra $C^*(\Gm)$. (Cf.\
\ci{Ren} for information
on groupoid  $C^*$-algebras; in the present case, $C^*(\Gm)$ is canonically
defined without
reference to a left Haar system, since $\Gm$ is a manifold and one may use
half-densities rather than
functions as elements of the algebra. Alternatively, the same algebra may be
defined with respect to
a left Haar system, each of whose measures is equivalent to the Lebesgue
measure in any local
co-ordinate system on the relevant fiber, which is a manifold. For convenience
we will choose the
latter option.)  The quantization of $\cin(\h^*)$ by $C^*(H)$ is a special case
of the groupoid
situation, as is the quantization of $\cin(T^*M)$ by the $C^*$-algebra of
compact operators on
$L^2(M)$, with a strict deformation quantization of Weyl type given in
\ci{NPLstr}.

We use the following notation: the base space of $\Gm$ is called $B$, the
source and target
projections are $s$ and $t$, respectively, and the left Haar system consists of
measures $\mu_b$ on
$t^{-1}(b)$, $b\in B$. The convolution product on $C^*(\Gm)$ is given (firstly
on $C_c(\Gm)$) by
\be
f*g(x)=\int_{t^{-1}(s(x))}d\mu_{s(x)}(y)\, f(xy)g(y^{-1}), \ll{3.16}
 \ee
and the involution is
\be
f^*(x)=\ovl{f(x^{-1})}. \ll{3.17}
\ee

 We assume that a right-representation $\pi^-$ of $C^*(\Gm)$ on a Hilbert space
$\H$ is given.
By a theorem of Renault \ci[II.1.21]{Ren}, this representation corresponds to a
representation
$\pi$ of $\Gm$ itself on $\H$ (to apply this theorem, we need to assume that
$\Gm$ is 2nd countable;
the other assumptions stated in \ci{Ren} are automatically satisfied for Lie
groupoids). Thus there
is a measure $\nu$ on $B$, and a Hilbert space $\H_b$ for ($\nu$-almost) every
$b\in B$, so that
$\H=\int_B^{\oplus}d\nu(b)\H_b$. The representative $\pi(x)$ of $x\in\Gm$ is
then a unitary map from
$\H_{s(x)}$ to $\H_{t(x)}$; note that $\pi(x)$ is not defined as an operator on
$\H$. Assuming that
$\Gm$ with given left Haar system is unimodular in the sense of \ci[I.3]{Ren}
(this assumption is
satisfied in all examples \ci{NPLstr,NPLcq}), the right-representation $\pi^-$
is given on $f\in
C_c(\Gm)$ by
 \be
(\pi^-(f)\ps)(b)=\int_{t^{-1}(b)} d\mu_b(y) f(y^{-1})\pi(y)\ps(s(y)). \ll{3.18}
\ee
Using (\ref{3.16}) and the left-invariance of the Haar system (which means that
$\mu_{s(x)}(E)=\mu_{t(x)}(xE)$ for each Borel set $E\subset t^{-1}\circ s(x)$)
it indeed follows
that $\pi^-(f)\pi^-(g)=\pi^-(g*f)$. We now define the rigging map on an
appropriate subspace
$L\subset \H$  by
\be
\la\ps,\phv\ra_{C_c(\Gm)}: x\raw  (\pi(x)\phv(s(x)),\ps(t(x)))_{t(x)},
\ll{3.19}
\ee
where the inner product on the right-hand side is the one in $\H_{t(x)}$.
Clearly, the rigging map (\ref{3.6}) is a special case of (\ref{3.19}).
By $L$ being `appropriate' we simply mean that it be chosen such that the
rigging map indeed takes
values in $C_c(\Gm)$. Checking the properties 1-4 stated at the beginning of
subsect.\ 3.1 is an easy
matter, given (\ref{3.16}) - (\ref{3.19}); one only needs the properties
$\pi(x)\pi(y)=\pi(xy)$,
$s(xy)=s(y),\,  t(xy)=t(x)$, and $s(x^{-1})=t(x),\, t(x^{-1})=s(x)$. Of course,
the algebra $\A$
should be contained in the commutant of $\pi^-(C^*(\Gm))$.
\begin{prop}
The rigging map (\ref{3.19}) is positive if $\Gm$ is amenable. \ll{posoid}
\end{prop}
{\em Proof.} The notion of amenability of a groupoid is defined in
\ci[II.3]{Ren}. We can simply copy
the proof of Proposition \ref{rmp}, the functions $g_j$ being given by the
functions $f_i$ of
Definition II.3.1 of \ci{Ren}. \Box
\section{Some examples}\eo
\subsection{Co-adjoint orbits and unitary representations of semidirect
products}
An important special case of symplectic reduction arises when one reduces
$T^*G$ with respect to the
right-action of a subgroup $H\subset G$; as we mentioned in the Introduction,
it was already pointed
out in \ci{KKS,GS,Wei87} that this reduction is the classical analogue of
Mackey's construction of
induced group representations (which in itself is a special case of Rieffel
induction, cf.\
\ci{Rie74,FD}, or the end of subsect.\ 3.2 above).
As a neat illustration of the general analogy between symplectic reduction and
Hilbert space
induction, we will now spell out how the representation theory of semidirect
product Lie groups of
the type $G=L\ltimes V$, with $V$ abelian, may be seen in this light. By the
Mackey theory
\ci{Mac,Var}, all unitary irreducible representations of $G$ are induced from
subgroups of the type
$H=S\ltimes V$, where $S\subset L$ is the stability group of a point $\pot\in
V^*$ under the dual
action of $L$. If $\pi_{\sg}$ is a  unitary irreducible representations of $S$,
one then induces
from representations $\pi_{(\sg,\pot)}$ defined by
$\pi_{(\sg,\pot)}(s,v)=\exp(i\la
\pot,v\ra)\pi_{\sg}(s)$.

The classical counterpart of this result of Wigner and Mackey would be that all
co-adjoint orbits in
$\g^*$ are (symplectomorphic to) Marsden-Weinstein reduced spaces of the form
$(T^*G)^{\O}\equiv J^{-1}(\O)/H$, with $H$ as above,
$\O\equiv \O_{(\sg,\pot)}=\O_{\sg}\oplus \pot$ a co-adjoint orbit in $\h^*={\bf
s}^*\oplus V^*$, and
$\O_{\sg}$ a co-adjoint orbit in ${\bf s}^*$. Here $J\equiv J_R :T^*G\raw
(\h^*)^-$ is the moment map
derived from the pull-back of the right-action of $H$ on $G$, cf.\ subsect.\
2.3 (par.\ following
Theorem \ref{tis}), whose notation and results we will freely use below. We
will now verify that this
is indeed the case.

Firstly, we need to check that $\O_{(\sg,\pot)}$ (which we again will simply
call $\O$ in what
follows)  as defined above is indeed a co-adjoint orbit of $H$; this follows
from the explicit action
of a semidirect product group on its dual, given in \ci[I.19]{GS}. Secondly, we
must demonstrate that
the map $J_L^{\O}: (T^*G)^{\O
}\raw
\g^*$, which we already know to be symplectic and equivariant (intertwining the
left-action of $G$
on $(T^*G)^{\O}$ and the co-adjoint action on $\g^*$), is injective (so that
$(T^*G)^{\O}$ is symplectomorphic to its image under $J_L^{\O}$), and
thirdly, it should follow that any orbit in $\g^*$ is such an image for
appropriately chosen $H$ and
$(\sg,\pot)$.

Using the left-trivialization of $T^*G$, we have seen that
$J_L^{\O}([x,p]_H)=xp$; putting
  $x=(l,v)\in G$, and $p=(\th,\pot)\in g^*$, where $\th\in {\bf l}^*$ is such
that $\th\upharpoonright {\bf s}$ lies in $\O_{\sg}$, we find
$J_L^{\O}([(l,v),(\th,\pot)]_H)=l\th+l\pot$. Here the right-hand side was
calculated using the
formula for the co-adjoint action of $G$ given in \ci[I.19]{GS}, and (following
this ref.) we have
written the co-adjoint action of $l$ on $\th$ simply as $l\th$, and the dual
action of $l$ on
$\pot\in V^*$ as $l\pot$. Now use the fact that the right-action $\rh$ of
$(s,w)\in H=S\ltimes V$ on
the point $((l,v),(\th,\pot))\in T^*G\simeq G\times \G^*$ is given by
$$ \rh_{(s,w)}((l,v),(\th,\pot))=((ls^{-1},v-ls^{-1}w),(s\th,\pot))$$
to conclude that the map $J_L^{\O}$ is indeed well-defined and injective on the
quotient
of $J^{-1}(\O)\subset T^*G$ by $H$. Finally, the fact that any co-adjoint orbit
in
$\g^*$ is obtained in this way follows from the classification of these orbits
in \ci[I.19]{GS}.

This result is closely related to a theorem in \ci{MRW}, which states that each
co-adjoint orbit in
$\g^*$ is symplectomorphic to a symplectic leaf in $(T^*L)/S$ for suitable $S$,
which $S$ is
exactly what we  used above. In addition, we mention the work of Rawnsley
\ci{Raw}, who related the
Wigner-Mackey representation theory of semidirect products  to the geometric
quantization of
certain of their co-adjoint orbits. This is quite different in spirit from our
approach, which in
this situation does not use any explicit correspondence between co-adjoint
orbits and irreducible
unitary representations (let alone geometric quantization), but rather
emphasizes the fact that both
are obtained by an induction procedure, which even employs the same class of
subgroups $H$  in the
classical and the quantum case. Moreover, even leaving quantum representation
theory aside, the
results of this subsection and the next, taken together with Corollary
\ref{mwis}, considerably
simplify the study of of actions of semidirect product or nilpotent Lie groups
on symplectic
manifolds.
 \subsection{Co-adjoint orbits and unitary representations
of nilpotent Lie groups}
A similar result holds when $G$ is nilpotent. Assuming $G$ to be connected
and simply connected for simplicity, the Dixmier-Kirillov theory \ci{CG}
establishes a bijective
correspondence between the co-adjoint orbits of $G$ and its irreducible unitary
representations. For
us, the main point is that all unitary representations are obtained by Mackey
induction from certain
subgroups $H$, and this inspires us to demonstrate that all co-adjoint orbits
of $G$ are
Marsden-Weinstein reduced spaces induced by the same $H$'s.

Pick a point $\pot\in\g^*$, and take $G_0\subset G$ the stability group of
$\pot$ under the
co-adjoint action. The essential implication of the nilpotence of $G$ is the
existence of a
so-called polarizing subalgebra $\h$, where $\g_0\subseteq\h\subseteq\g$, and
$\la\pot,[X,Y]\ra=0$ for all $X,Y$ in $H$. With $H=\exp\h$, one then induces
from the representation
$\pi_{\pot}(H)$ given by $\pi_{\pot}(\exp(X))=\exp(i\la\pot,X\ra)$.
Representations thus obtained are
unitarily equivalent iff the various $\pot$ one starts from lie in the same
orbit, and all
 irreducible unitary representations of $G$ are obtained in this way.

To find the classical analogues of these statements, we first notice that $\h$
being polarizing
relative to $\pot$ simply means that $\pot_r\equiv\pot\upharpoonright\h\in\h^*
$ is stable under the
co-adjoint action of $H$. Hence $\pot_r$ is a co-adjoint orbit in $\h^*$, and
we are done if we can
show that $G\pot\, \simeq J^{-1}(\pot_r)/H$ as symplectic spaces (here $J\equiv
J_R:T^*G\raw
(\h^*)^-$, as in the previous subsection). This is indeed the case.

First, note that $J^{-1}(\pot_r)=G\times (H\pot)$ (in the left-trivialization
of $T^*G\simeq
G\times \g^*$). To prove this,
observe that the set $\Sigma=\{p\in\g^*|p \upharpoonright \h=\pot_r\}\subset
\g^*$ is a copy of $\Bbb R^n$ in $\g^*$, with  $n={\rm dim}\, \g-{\rm
dim}\,\h$.
On the other hand, $H$, being connected, simply connected, and nilpotent, acts
unipotently on $\g^*$,
so that by \ci[Cor.\ 3.1.5]{CG} the orbit $H\pot$ is homeomorphic to $\Bbb
R^m$, with $m={\rm dim}\,
H-{\rm dim}\, G_0$. But $H$ is a polarizing subgroup of $G$, hence $n=m$ by
\ci[Thm.\ 1.3.3]{CG}.
 Secondly, if $p=h\pot$ for some $h\in H$ then $p\upharpoonright
\h=\pot\upharpoonright
\h=\pot_r$, since the map $p\raw p\upharpoonright\h$ intertwines the co-adjoint
action of $H$ on
$\g^*$ with its action on $\h^*$. Hence $G\times (H\pot)\subseteq
J^{-1}(\pot_r)$. The claim follows.

As $H\pot\, =H/G_0$, we have $J^{-1}(\pot_r)/H=(G\times (H/G_0))/H$, with the
right H-action
$\rh$ defining the quotient given by $\rh_h(x,p)=(xh,h^{-1}p)$. Hence $(G\times
(H/G_0))/H\simeq
G/G_0=G\pot$, and this is a symplectomorphism implemented by  the map
$J_L^{\O}:J^{-1}(\O)/H\raw \g^*$ (cf.\ the previous subsection), with
$\O=\pot_r$. Since $\pot$, and
hence the co-adjoint orbit $G\pot$, was arbitrary, we have indeed established
that any co-adjoint
orbit in a connected and simply connected nilpotent Lie group is obtained by
Marsden-Weinstein
reduction from a polarizing subgroup and a zero-dimensional orbit. This
establishes a perfect correspondence between classical and quantum induction in
this case.
\subsection{The generalized Yang-Mills construction}
Let $(P,H,Q,pr')$ be a principal fiber bundle with connected compact gauge
group $H$ and projection
$pr':P\raw Q=P/H$; we assume $P$ connected as well. Then $H$ acts from the
right on $T^*P$ by
pull-back with moment map $J$, and we have a full dual pair
$(T^*P)/H\stackrel{pr}{\law}T^*P\stackrel{J}{\raw}(\h^*)^-$, where $pr$ is the
canonical projection
onto the given quotient space \ci{Wei83}. With extra assumptions on simple
connectedness, one even
obtains a classical equivalence bimodule, so that $\h^*$ and $(T^*P)/H$ are
Morita equivalent Poisson
manifolds with $T^*P$ as their equivalence bimodule, cf. \ci{Xu91} or
Definition \ref{ceb} in
subsect.\ 2.1 above. However, by the argument in \ci[sect.\ 8]{Wei83}, there is
a bijective
correspondence between the symplectic leaves in $(T^*P)/H$ and $\h^*$, which is
given explicitly in
\ci{GS,Mar}. There it is shown that the leaves of $(T^*P)/H$ are fiber bundles
over $T^*Q$ with
a co-adjoint orbit in $\h^*$ as fiber. This suggests that   $(T^*P)/H$ and
$\h^*$ are
Morita-equivalent without any further assumption; their equivalence bimodule
may be different from
$T^*P$ in general.

 In any case, the essential point is that $pr^*\cin((T^*P)/H)$ Poisson commutes
with $J^*\cin(\h^*)$ in $\cin(T^*P)$, so that,  starting from any given
realization
$\rh:S_{\rh}\raw\h^*$, we obtain an induced representation
$\pi^{\rh}:\cin((T^*P)/H)\raw
\cin(S^{\rh})$
 by the construction in subsect.\ 2.1
(or, equivalently, we find a Poisson map $J^{\rh}:S^{\rh}\raw (T^*P)/H$ for
each such $\rh$).
If we take $S_{\rh}$ to be a co-adjoint orbit in $\h^*$ then $S^{\rh}$ is a
symplectic leaf of
$(T^*P)/H$, which plays the role of the phase space of a particle in a
Yang-Mills field, as
originally observed by Sternberg (cf.\  \ci{Wei78,GS,Mar} for a comprehensive
discussion), whence
the name `Yang-Mills construction'. Inducing from an arbitrary realization
$S_{\rh}$ leads to the
`generalized Yang-Mills construction' \ci{Wei87,Xu92}.

The Yang-Mills construction was quantized in \ci{NPLstr}, where we exploited
the fact that
$\cin((T^*P)/H)$ is the Poisson algebra canonically associated to the Lie
algebroid $(TP)/H$
\ci{CDW}. Here we wish to briefly give a general construction based on Rieffel
induction.
Namely, the quantum analogue of the full dual pair mentioned above is  the
imprimitivity
bimodule $\K(L^2(P))^H\raw L^2(P)\law C^*(H)$, which involves the $H$-invariant
compact operators on
$L^2(P)$. To see this, one may start from the right-action $\pi^-$ of
$C_c(H)\subset C^*(H)$ on
$C_c(P)\subset L^2(P)$, provided (via (\ref{3.5})) by the unitary
representation $\pi$ of $H$ on
$L^2(P)$, which comes from the right-action  defining the principal bundle.
For simplicity, we put
an $H$-invariant measure $\mu$ on $P$ (always possible, as $H$ is compact),
which defines $L^2(P)$.
Then $(\pi(h)\ps)(x)=\psi(xh)$.

The rigging map into $C^*(H)$ is given by (\ref{3.6}), and is
positive by Lemma \ref{pos}. Using the fact that the $C^*$-norm $\n f\n$ of
$f\in C_c(H)$ is dominated
by its $L^1$-norm, as well as the Cauchy-Schwartz inequality and $\int_H
dh\,=1$, one finds that
$\n \la\ps,\phv\ra_{C_c(H)} \n\, \leq \; \n\ps\n_2\, \n\phv\n_2$, so that one
can extend (\ref{3.6})
by continuity to a rigging map defined on $\H$ with values in $C^*(H)$.
Note that, since $H$ acts freely on $P$, we can choose a (discontinuous)
cross-section $s:Q\raw P$,
which leads to a natural isomorphism $L^2(P)\simeq L^2(Q)\ot L^2(H)$, where
$\pi(H)$ acts trivially
on the first factor and via the right-regular representation on the second one.
Since the rigging
map then amounts to convolution over  $H$ (times an inner product in $L^2(Q)$),
this implies that
the image of the rigging map defined on $C_c(P)$ is dense in $C_c(H)$, hence in
$C^*(H)$.

Now consider the imprimitivity algebra \ci{Rie74,FD} (also cf.\ subsect.\ 3.1)
$\A$ defined by
$\H=L^2(P)$, $\B=C^*(H)$, and the rigging map (\ref{3.6}).
 $\A$ is generated by operators of the form $T_{(\ps,\phv)}$, whose action on
$\zt\in L$ is defined
by $T_{(\ps,\phv)}\zt= \pi^-(\la\phv,\zt\ra_{C^*(H)})\ps$. Starting with
$\ps,\phv,\zt\in C_c(P)$,
and using (\ref{3.5}), (\ref{3.6}), we find that $T_{(\ps,\phv)}$ is
Hilbert-Schmidt with kernel
given by $K_{(\ps,\phv)}(x,y)=\int_H dh\, \ps(xh)\ovl{\phv(yh)}$. From the
property
$K(xh,yh)=K(x,y)$ for all $h\in H$ and $x,y\in P$ we infer that
$T_{(\ps,\phv)}$ commutes with all
$\pi(h)$. Hence the $C^*$-algebra generated by these operators is clearly $\A=
\K(L^2(P))^H$.
The $\A$-rigging map is defined by $\mbox{}_{\A}\la\ps,\phv\ra=T_{(\ps,\phv)}$,
and all relevant
conditions  are now automatically satisfied (cf.\ \ci[sect.\ 6]{Rie74}) for
$L^2(P)$ to become a
$\K(L^2(P))^H\, -\, C^*(H)$ imprimitivity bimodule.

Physically, $\A$ is the `universal algebra of observables' of a particle in a
Yang-Mills field with
gauge group $H$ \ci{NPLstr}, and is the quantum counterpart of the Poisson
algebra $\cin((T^*P)/H$,
which plays this role in classical mechanics \ci{Wei78}. By the Rieffel
imprimitivity theorem
\ci{Rie74} (also cf.\ subsect.\ 3.1 above) combined with the strong Morita
equivalence between
$\A$ and $\B=C^*(H)$ established above, all its representations are induced by
representations of
$C^*(H)$, hence by unitary representations of $H$. The explicit form of these
induced representations
is then given by the Moscovici induction technique discussed at the end of
subsect.\ 3.2 above as a
special case of the Rieffel process. Starting from a unitary representation
$\plg(H)$ on a Hilbert
space $\hlg$, one finds that the Hilbert space $\til{\H}^{\ch}$ carrying the
induced representation
$\til{\pug}(\A)$ is just the $L^2$-closure of the space $\Gm_{\ch}$ of smooth
compactly supported
cross-sections  of the vector bundle $P\times_H\hlg$ associated to the
principal bundle
$(P,H,Q,pr')$. This realization was previously found by different means
\ci{NPLstr}; we note that
the space $\Gm_{\ch}$ is a useful domain of essential self-adjointness of
various unbounded
operators of physical relevance.
\subsection{The illusion of time}
The classical relativistic particle in Minkowski space-time is discussed in an
elegant covariant
symplectic formalism in \ci{GS}. As we failed to find a convincing quantization
of this approach in
the literature, we here discuss this system using  Rieffel induction.

The classical setup consists of the cotangent bundle $S=T^*\Bbb R^4$ and the
group $H=\Bbb R$, which
acts on $S$ by generating geodesic motion on the flat space-time $\Bbb R^4$. If
we write
$(x(\ta),p(\ta))$ for the result of the action of $\ta\in\Bbb R$ on $(x,p)\in
S$, we thus have
\be
(x^{\mu}(\ta),p_{\nu}(\ta))=(x^{\mu}+p^{\mu}\ta,p_{\nu}), \ll{rp1} \ee
 where
$p^{\mu}=g^{\mu\nu}p_{\nu}$, with $g^{\mu\nu}$ the metric diag$(1,-1,-1,-1)$.
If the symplectic form
is taken to be $dx^{\mu}\wedge dp_{\mu}$, this action corresponds to the moment
map $J:T^*\Bbb
R^4\raw \h^*=\Bbb R$ defined by $J(x,p)=g^{\mu\nu}p_{\mu}p_{\nu}/2$. The
observables on $S$ are the
functions $f\in\cin(S)$ which Poisson-commute with $J$, that is, satisfy
$p^{\mu}\partial f/\partial
x^{\mu}=0$. We now reduce $S$ with respect to the co-adjoint orbit $S_{\rh}=
\{m^2/2\}\in\h^*$, and
find that the reduced phase space $S^{\rh}$ consists of two disconnected
copies, one with $p_0>0$
and one with $p_0<0$; the latter may consistently be ignored by imposing the
additional constraint
$p_0>0$. Each copy consists of
equivalence classes of points in $S*_{\Bbb R} \{m^2/2\}\equiv \{(x,p)\in
S|p^2=m^2\}$, where points
are in the same equivalence class iff they are connected by the flow
(\ref{rp1}). Therefore, a point
in the physical (i.e., reduced) phase space, identified with a physical state
of the relativistic
particle, consists of an entire particle trajectory through space-time.

Using the prescription proposed in this paper, it is completely straightforward
to quantize this
model.
We take $\H=L^2(\Bbb R^4)$ (regarded as functions on space-time), which carries
a representation of
$H=\Bbb R$ given by $\pi(\ta)\ps=\exp(i\ta\Box)\ps$ (with
$\Box=g^{\mu\nu}\partial/\partial x^{\mu}
\partial/\partial x^{\nu}$). We define the rigging map (\ref{3.6}) on
$L=\cci(\Bbb R^4)$,
and induce from the irreducible unitary  representation $\pi_{m^2}: \ta\raw
\exp(i\ta m^2)$ on
$\H_m=\Bbb C$.  With (\ref{3.1}), we find that the form $(\cdot,\cdot)_0$ of
the Rieffel induction
process is given by (note that $L\ot\hlg=L$ in the present case)
\be
(\ps,\phv)_0=\int_{\Bbb R}d\ta\, e^{i\ta m^2}\int_{\Bbb R^4} d^4x\,
(e^{i\ta\Box}\ps)(x)\ovl{\ph(x)}
 =\int_{\Bbb R^4} \frac{d^4
p}{(2\pi)^3}\dl(p^2-m^2)\hat{\ps}(p)\ovl{\hat{\phv}(p)}. \ll{rp2}
\ee
Hence the final representation space $\H^{m^2}$ consists of solutions $\ps$ of
the Klein-Gordon
equation $(\Box+m^2)\ps=0$, with either positive or negative energy $p_0$,
whose Fourier transforms
are square-integrable with respect to the measure $d^3p/p_0$, which one finds
by integrating the delta
function in (\ref{rp2}). Alternatively, one may follow the construction of
$\til{\H}^{\ch}$
explained after (\ref{3.4}), with $\ch=m^2$, and arrive at the same result. The
quantum
observables are those bounded operators which commute with the multiplication
operator $p^2$.

The interpretation of the quantum states is similar to the classical ones: each
vector in
$\H^{m^2}$ consists of a wave function on space-time. The propagation of states
in time, familiar
from non-relativistic mechanics, here has to be derived from external
considerations.
\subsection{Finite $W$-algebras}
Inspired by developments in conformal field theory and integrable systems, the
concept of a finite
$W$-algebra was recently introduced \ci{BT}. This subject provides an
illustration of Rieffel
induction applied to an algebra of unbounded operators, and it appears to us
that our quantization
method applied here is simpler than the BRST and Lie algebra cohomology
techniques
used in \ci{BT}.

The setting is a  Lie group $G$ with  Lie subgroup $H$. In the context of
$W$-algebras, $G$ is
semi-simple and $H$ is nilpotent, but these assumptions hardly play a role in
our discussion.  $H$
acts on $\g^*$ by restriction of the co-adjoint representation. This action
preserves the Lie-Poisson
structure, and the corresponding generalized moment map $j$ is simply given by
$j(\th)=\th
\upharpoonright \h$. Picking an orbit $\O\subset\h^*$, we can define the
Poisson reduced space
$j^{-1}(\O)/H$ \ci{MR}, and the corresponding classical finite $W$-algebra is
the space of real
polynomials  $W_c(G,H,\O)={\rm pol}_{\Bbb R}[j^{-1}(\O)/H]$, equipped with the
reduced Poisson
structure.

To quantize, it is convenient to have an equivalent definition at hand.
Recall \ci{AM} that $G\backslash T^*G\simeq \g^*$, so that
$\mbox{}^G\cin(T^*G)$
(the space of left $G$-invariant smooth functions on $T^*G$)
is Poisson-isomorphic to the
Lie-Poisson algebra $\cin(\g^*)$, whose subspace of polynomials ${\rm
pol}_{\Bbb R}[\g^*]$ is well
defined. The right $H$-action on $T^*G$ quotients to the co-adjoint action on
$\g^*$. Hence the space
$\mbox{}^G\cin(T^*G)^H$ may be restricted to the space $A\subset\cin(T^*G)$ of
$H$-invariant polynomials on $\g^*$. $A$ is a Poisson algebra with the
canonical Poisson bracket on
$T^*G$, which is left and right $G$-invriant, and therefore can  consistently
be restricted to $A$.
If we now take $S=T^*G$, $S_{\rh}=\O$ a co-adjoint orbit in $\h^*$, and
$\rh=i_{\O}$ the injection of
$\O$ in $\h^*$, we obtain an induced representation $\pi^{\O}(A)$ on the
Marsden-Weinstein reduced space  $S^{\O}=J^{-1}(\O)/H$, cf.\ subsect.\ 2.1 (we
here write $\pi^{\O}$
etc.\ for  $\pi^{i_{\O}}$). As before, $J:T^*G\raw (\h^*)^-$ is the moment map
coming from the right-action of $H$ on $T^*G$. It is easily seen that
$\pi^{\O}(A)\simeq W_c(G,H,\O)$
as Poisson algebras.

In this formulation, quantization is a piece of cake. Firstly, the quantization
of the Lie-Poisson
algebra ${\rm pol}_{\Bbb R}[\g^*]$ is the operator algebra ${\cal U}(\g)_{\rm
sa}$, which consists of
the symmetric elements of the universal enveloping algebra of $G$ (hence the
quantization of the
complexified Poisson algebra ${\rm pol}_{\Bbb C}[\g^*]$ is ${\cal U}(\g)$
itself). We here regard
${\cal U}(\g)$ as the  ${\rm Op}^*$-algebra of left-invariant differential
operators defined on the
common dense domain  $\cci(G)\subset L^2(G)$ \ci{Sch} (to be compared with
${\rm pol}_{\Bbb
C}[\g^*]$ being the left-invariant polynomials on $T^*G$). This quantization is
just the
infinitesimal and unbounded version of Rieffel's deformation quantization of
$C_0(\g^*)$ by the
group  algebra  $C^*(G)$ \ci{Rie89a}.

To see the connection between the two, we start by taking a
smooth function $f$ on $\g^*$ whose Fourier transform $\check{f}$ has compact
support on $\g$. Thus a
function $\til{f}$ may be defined for sufficiently small $\pl$ on a
neighbourhood  of the identity of
$G$,  by $\til{f}(\exp(-\pl X))=\pl^{-n}\check{f}(X)$ (with $n=\dim G$). Then
$\til{f}$ becomes a
smooth function on $G$ by putting it equal to zero elsewhere.  We then define
the deformation
quantization $Q_{\pl}(f)$  as an operator on $L^2(G)$ by
$Q_{\pl}(f)=\pi_R(\til{f})$, where $\pi_R$
is the right-regular representation of $G$ and $C^*(G)$ on $L^2(G)$ (hence
$(Q_{\pl}(f)\ps)(x)=\int_G
dy\, \til{f}(y)\ps(xy)$ for $\ps\in L^2(G)$). If we restrict $\ps$ to lie in
$\cci(G)$, then a short
formal computation shows that this quantization may be extended to any
$f\in{\rm pol}_{\Bbb
C}[\g^*]$, and  that the final result is defined for arbitrary  values of
$\pl$. Explicitly, one
finds that a monomial $\til{X}_1\ldots \til{X}_l\in {\rm pol}_{\Bbb R}[\g^*]$
(where, as before,
$\til{X}\in\g\subset \cin(\g^*)$ is defined by $\til{X}(\th)=\la\th,X\ra$) is
quantized by $(-i\pl)^l
\lm(X_1\ldots X_l)\in {\cal S}(\g)$. Here $\lm$ is symmetrization, and this
quantization is
identified with the corresponding element in the right-regular representation,
as before.

We now follow the Rieffel induction process with $L=\cci(G)$, $\A={\cal
U}(\g)^H$ (consisting of
invariants under the adjoint action of $H$), and $\B=\cci(H)$. $\A$ acts on $L$
as indicated above,
and $\B$ acts on $L$ by $\pi^-(f)\ps=\ps*f$ (convolution over $H$). We now
exploit the fact that $H$
is nilpotent, which implies that there is an irreducible  unitary
representation $\pi_{\O}$ on a
Hilbert space $\H_{\O}$  defined by the orbit $\O$ \ci{CG} (if $H$ is not
simply connected, this holds
provided that the orbit satisfies a suitable integrality condition).
The quantum $W$-algebra $W_q(G,H,\O)$ is then simply the induced representation
$\pi_q^{\O}(\A)$,
which is an algebra of unbounded operators acting on the dense domain
$D=\cci(G)\til{\ot}\H_{\O}\subset \H^{\O}$ (see (\ref{3.2}), with $\ch=\O$).
Explicitly, $\H^{\O}$ is
of course just the representation space obtained by inducing $\pi_{\O}(H)$ to
$\pi^{\O}(G)$ by the
Mackey procedure.

In the Blattner realization $\til{\H}^{\O}$ of $H$-equivariant functions
$\ps:G\raw\H_{\O}$ (that is, $\til{\ps}(xh)=\pi_{\O}(h^{-1}\til{\ps}(x)$) for
which
 $(\til{\ps},\til{\ps})_{\O}$ (inner
product in $\H_{\O}$) is square-integrable on $G/H$, the corresponding domain
$\til{D}$ consists of
those functions in $\til{\H}^{\O}$ which are smooth and the projection of whose
support on $G$ onto
$G/H$ is compact.
 Note, that $d\til{\pi}^{\O}({\cal U}(\g))$ acts on $\til{\ps}\in
\til{\H}^{\O}$ by hitting the
argument of $\til{\ps}$ from the left (e.g.,
$d\til{\pi}^{\O}(X)\til{\ps}(x)=d/dt\,
\til{\ps}(\exp(tX)x)_{|t=0}$ for $X\in\g$), which trivially preserves
$H$-equivariance of
$\til{\ps}$,
whereas $\til{\pi}_q^{\O}$ maps ${\cal U}(\g)^H$ into differential operators
hitting this
argument from the right. This still preserves the equivariance  on account of
the $H$-invariance of
elements of $\A$.
\subsection{Reduction by a groupoid algebra}
Our final example probably provides the simplest illustration of the use of
groupoids in constrained
systems. The classical system has $S=T^*\Bbb R^m$, and the aim is to eliminate
one degree of
freedom. This may be done by imposing the single constraint $p_1=0$, and reduce
with respect to the
corresponding action  of $H=\Bbb  R$ on $S$. However, it is more instructive to
start from a
Poisson map $J: T^*\Bbb R^m\raw (T^*\Bbb R)^-$. The observables have to commute
with the constraints
$J^*\cin(T^*\RE)$, and are just the functions which do not depend on $x^1,p_1$.
 With $S_{\rh}=
T^*\Bbb R$ and $\rh$ the identity map, the reduction procedure of subsect.\ 2.1
then
painlessly leads to the reduced phase space $S=T^*\Bbb R^{m-1}$, with the
obvious action of the
observables.

 Since $\cin(T^*\Bbb R^n)$ is the Poisson algebra defined by the Lie algebroid
$T\Bbb R^n$, its
quantization is the groupoid algebra $C^*(\RE^n\times\RE^n)=\K(L^2(\RE^n))$
\ci{NPLstr} (also cf.\
subsect.\ 3.4 above). Therefore, taking $n=m$, the quantization of the
unconstrained system is given
by  the defining representation of $\K(L^2(\RE^m))$ on $\H=L^2(\RE^m)$, and the
quantum algebra of
the constraints is (put $n=1$) $\K(L^2(\RE))$. To use the procedure of
subsect.\ 3.4, we identify
$\H$ as the direct integral $\int_{\RE}^{\oplus} dx\, \H(x)=L^2(\RE)\ot
L^2(\RE^{m-1})$, with
$\H(x)=L^2(\RE^{m-1})$. It is convenient to work with suitable dense subspaces,
so we take
$L=C_c(\RE^m)$, $\A= \Bbb I\ot \K_c(L^2(\RE^{m-1}))$, and $\B=\K_c(L^2(\RE))$.
Here
$\K_c(L^2(\RE^n))$ consists of the Hilbert-Schmidt operators whose kernel is in
$C_c(\RE^n,\RE^n)$.
If we identify a Hilbert-Schmidt operator $f$ with its kernel, then the
representation (\ref{3.18})
reads \be
(\pi^-(f)\ps)(x^1,\ldots,x^m)=\int_{\RE} dx\, f(x,x^1)\ps(x,x^2,\ldots,x^m).
\ll{gr1}
\ee
The rigging map (\ref{3.19}) is
\be
\la\ps,\phv\rab:(x,y)\raw \int_{\RE^{m-1}}dx^2\ldots dx^m\,
\phv(y,x^2,\ldots,x^m)
\ovl{\ps(x,x^2,\ldots,x^m)}. \ll{gr2}
\ee
We now induce from the identity representation $\pi_{\rm id}$ of $\B$ on
$\H_{\rm id}=L^2(\RE)$.
The space $L\ot \H_{\rm id}$ may be identified with a space of functions in
$m+1$ variables, so that
the form (\ref{3.1}) becomes
\be
(\Psi,\Phi)_0=\int_{\RE^{m+1}} dx^0\ldots dx^m\,
\Psi(x^0,x^0,x^2,\ldots,x^m) \ovl{\Phi(x^1,x^1,x^2,\ldots,x^m)}. \ll{gr3}
\ee
In particular, $(\Psi,\Psi)_0=\int dx^2\ldots dx^m\, |\int dx\,
\Psi(x,x,x^2,\ldots,x^m)|^2$, so
that $(\cdot,\cdot)_0$ is positive semi-definite, as it should be by
Proposition \ref{posoid}.
The closure $\H^{\rm id}$ of the quotient of $L\ot\H_{\rm id}$ by the null
space $\H_0$ of
$(\cdot,\cdot)_0$ is naturally realized as $L^2(\RE^{m-1})$: if we define
$U:L\ot \H_{\rm id}\raw
L^2(\RE^{m-1})$ by $(U\Psi)(x^2,\ldots,x^m)=\int dx \,
\Psi(x,x,x^2,\ldots,x^m)$ then $U$ exactly
annihilates $\H_0$, and quotients to a unitary map $\til{U}$ from $\H^{\rm id}$
to  $L^2(\RE^{m-1})$.
The corresponding representation $\til{U}\pi^{\rm id}\til{U}^{-1}$ of the
algebra of observables
$\A$ is simply the identity representation of $\K_c(L^2(\RE^{m-1}))$ on
$L^2(\RE^{m-1})$.

This result may be much ado about nothing, but we wish to point out that this
extremely simple
constrained  system  cannot be quantized by the BRST method without serious
{\em ad hoc}
modifications \ci{LL}.

 \end{document}